\documentclass[fleqn,usenatbib]{mnras}
\usepackage{pgfplotstable,filecontents}
\usepackage{newtxtext,newtxmath}
\usepackage{csvsimple}
\usepackage[normalem]{ulem}

\usepackage[T1]{fontenc}

\DeclareRobustCommand{\VAN}[3]{#2}
\let\VANthebibliography\thebibliography
\def\thebibliography{\DeclareRobustCommand{\VAN}[3]{##3}\VANthebibliography}

\usepackage{graphicx}	
\usepackage{amsmath}	

\newcommand{\kms}{km~s$^{-1}$}
\newcommand{\Msun}{M$_{\odot}$}
\newcommand{\Rc}{$R_{\mathrm{c}}$}
\newcommand{\rms}{$\sigma_{\mathrm{rms}}$}
\newcommand{\siggs}{$\sigma_{\mathrm{gs,los}}$}
\newcommand{\sigobs}{$\sigma_{\mathrm{obs,los}}$}
\newcommand{\Mgas}{$M_{\mathrm{gas}}$}
\newcommand{\Mvir}{$M_{\mathrm{vir}}$}

\newcommand{\surfgas}{$\Sigma_{\mathrm{gas}}$}
\newcommand{\omeobs}{$\omega_{\mathrm{obs}}$}
\newcommand{\rotaxis}{$\phi_{\mathrm{rot}}$}
\newcommand{\virialpha}{$\alpha_{\mathrm{vir}}$}
\newcommand{\Lco}{$L_{\mathrm{CO(2-1)}}$}
\newcommand{\Fco}{$F_{\mathrm{CO(2-1)}}$}
\newcommand{\CO}{$^{12}$CO(2-1)}

\newcommand{\punit}{K~cm$^{-3}$}

\title[WISDOM -- XV.\ GMCs in NGC~5806]{WISDOM Project -- XV.\ Giant Molecular Clouds in the Central Region of the Barred Spiral Galaxy NGC~5806}

\author[Choi et al.]{
  Woorak Choi,$^{1}$\thanks{E-mail: woorak.c@gmail.com}
  Lijie Liu,$^{2,3}$
  Martin Bureau,$^{4,5}$\thanks{E-mail: martin.bureau@physics.ox.ac.uk}
  Michele Cappellari,$^{4}$
  Timothy A. Davis,$^{6}$
  \newauthor
  Jindra Gensior,$^{7}$
  Fu-Heng Liang,$^{4}$
  Anan Lu,$^{8}$
  Thomas G. Williams,$^{4,9}$
  Aeree Chung$^{1}$\thanks{E-mail: achung@yonsei.ac.kr}
  \\
  $^{1}$Department of Astronomy, Yonsei University, 50 Yonsei-ro, Seodaemun-gu, Seoul 03722, Republic of Korea\\
  $^{2}$Cosmic Dawn Center (DAWN), Technical University of Denmark, DK2800 Kgs.\ Lyngby, Denmark \\
  $^{3}$DTU-Space, Technical University of Denmark, Elektrovej 327, DK2800 Kgs.\ Lyngby, Denmark\\
  $^{4}$Sub-department of Astrophysics, Department of Physics, University of Oxford, Keble Road, Oxford OX1 3RH, UK
  \\$^{5}$Yonsei Frontier Lab and Department of Astronomy, Yonsei University, 50 Yonsei-ro, Seodaemun-gu, Seoul 03722, Republic of Korea\\
  $^{6}$Cardiff Hub for Astrophysics Research and Technology, School of Physics and Astronomy, Cardiff University, Queens Buildings, Cardiff, CF24 3AA, UK\\
  $^{7}$Institute for Computational Science, Winterthurerstrasse 190, 8057 Z\"{u}rich, Switzerland\\
  $^{8}$McGill Space Institute and Department of Physics, McGill University, 3600 rue University, Montreal, QC H3A 2T8, Canada\\
  $^{9}$Max Planck Institut f\"{u}r Astronomie, K\"{o}nigstuhl 17, 69117 Heidelberg, Germany
}

\date{Accepted XXX. Received YYY; in original form ZZZ}

\pubyear{2022}

\begin{document}
\label{firstpage}
\pagerange{\pageref{firstpage}--\pageref{lastpage}}
\maketitle

\begin{abstract}
    We present high spatial resolution ($\approx24$~pc) Atacama Large Millimeter/sub-millimeter Array \CO\ observations of the central region of the nearby barred spiral galaxy NGC~5806. NGC~5806 has a highly structured molecular gas distribution with a clear nucleus, a nuclear ring and offset dust lanes. We identify $170$ spatially- and spectrally-resolved giant molecular clouds (GMCs). These clouds have comparable sizes ($R_{\mathrm{c}}$) and larger gas masses, observed linewidths ($\sigma_{\mathrm{obs,los}}$) and gas mass surface densities than those of clouds in the Milky Way disc. The size -- linewidth relation of the clouds is one of the steepest reported so far ($\sigma_{\mathrm{obs,los}}\propto R_{\mathrm{c}}^{1.20}$), the clouds are on average only marginally bound (with a mean virial parameter $\langle\alpha_{\mathrm{vir}}\rangle\approx2$), and high velocity dispersions are observed in the nuclear ring. These behaviours are likely due to bar-driven gas shocks and inflows along the offset dust lanes, and we infer an inflow velocity of $\approx120$~\kms\ and a total molecular gas mass inflow rate of $\approx5$~M$_\odot$~yr$^{-1}$ into the nuclear ring. The observed internal velocity gradients of the clouds are consistent with internal turbulence. The number of clouds in the nuclear ring decreases with azimuthal angle downstream from the dust lanes without clear variation of cloud properties. This is likely due to the estimated short lifetime of the clouds ($\approx6$~Myr), which appears to be mainly regulated by cloud-cloud collision and/or shear processes. Overall, it thus seems that the presence of the large-scale bar and gas inflows to the centre of NGC~5806 affect cloud properties.
\end{abstract}




\section{Introduction}
\label{sec:intro}

As giant molecular clouds (GMCs) are the gas reservoirs where all star formation occurs, elucidating their life cycles is crucial to understand the formation and evolution of galaxies. Early GMC studies were conducted only in our own Milky Way (MW) and Local Group galaxies such as the Large Magellanic Cloud (LMC; e.g.\ \citealt{fukui2008}), Small Magellanic Cloud (SMC; e.g.\ \citealt{muller2010}), M~31 \citep[e.g.][]{rosolowsky2007} and M~33 \citep[e.g.][]{rosolowsky2003,rosolowsky2007etal}, showing that GMCs in those galaxies have properties similar to each other and follow the same size -- linewidth relation (e.g.\ \citealt{larson1981,bolatto2008}). As the resolution and sensitivity of molecular line observations improved, GMC studies were extended to extragalactic objects, revealing deviations from the properties of Local Group galaxy GMCs (e.g.\ \citealt{bolatto2008,rosolowsky2021}). For instance, the cloud properties in some late-type galaxies (LTGs) vary depending on galactic environments and do not universally obey the usual scaling relations (e.g.\ M~51, \citealt{hughes2013,colombo2014m51}; NGC253, \citealt{leroy2015ngc253}). The first study of GMCs in an early-type galaxy (ETG; NGC~4526, \citealt{utomo2015}) revealed that the GMCs in that galaxy do not have a clear correlation between size and linewidth but are brighter, denser and have higher velocity dispersions than GMCs in the MW disc (MWd) and Local Group galaxies. On the other hand, \citet{liu2021ngc4429} recently reported that the GMCs in the ETG NGC~4429 have an unusually steep size -- linewidth relation. These results indicate that galactic environment affects GMC properties, so more GMCs studies in galaxies with different morphologies and substructures are required to quantify these variations and understand the physics behind them.

Barred disc galaxies are known to have gas streaming to their centres due to their non-axisymmetric gravitational potentials \citep[e.g.][]{sormani2015_bar_gasinflow}. Several CO surveys have reported higher central molecular gas mass concentrations in barred than non-barred disc galaxies \citep[e.g.][]{sakamoto1999_bar_concentration,sun2020_bar_phangs}. Recent high spatial resolution CO observations of barred disc galaxies have also shown that these objects possess several distinct structures mimicking those present at optical wavelengths (e.g.\ nuclear rings, bars and spiral arms), with non-circular motions \citep[e.g.][]{salak2016,BewketuBelete2021,sato2021}. Thus, barred disc galaxies allow to investigate the properties of GMCs (e.g.\ scaling relations) in different environments, particularly the bars themselves. Despite this, however, very few studies investigating GMCs in barred galaxies exist \citep[e.g.][]{hirota2018_m83_bar,maeda2020a_n1300_bar,sato2021}.

As part of the mm-Wave Interferometric Survey of Dark Object Masses (WISDOM) project, we analyse here the properties and dynamics of individual GMCs in the centre of the barred spiral galaxy NGC~5806 located in the field.  WISDOM aims to use the high angular resolution of Atacama Large Millimeter/sub-millimeter Array (ALMA) to study (1) the physical properties and dynamics of GMCs in the centres of galaxies and how these link to star formation \citep[e.g.][]{liu2021ngc4429,liu2022_ngc404,lu2022_wisdom} and (2) the masses of the supermassive black holes lurking at the centres of the same galaxies.
 
This paper is structured as follows. In \autoref{sec:data_identification}, we describe the data and methodology used to identify GMCs in NGC~5806. The cloud properties, their probability distribution functions and their mass distribution functions are discussed in \autoref{sec:properties}. In \autoref{sec:kinematics}, we investigate the kinematics of the clouds and their origins. In \autoref{sec:dynamics}, we assess the dynamical states and degrees of virialisation of the clouds. We further discuss the morphology and velocity dispersion of the molecular gas, the formation, destruction, scaling relations and virialisation of the GMCs, the clouds in the nuclear ring and the CO-to-H$_2$ conversion factor in \autoref{sec:discussion}. We summarise our findings in \autoref{sec:conclusions}.

\section{Data and Cloud Identification}
\label{sec:data_identification}

\subsection{Target}
\label{sec:target}

NGC~5806 is a nearby barred spiral galaxy (SAB(s)b) located at R.A.=$15^{\mathrm{h}}00^{\mathrm{m}}00\fs5$, Dec.$=1\degr53\arcmin30\arcsec$ (J2000). Throughout this paper, we adopt a distance $D=21.4$~Mpc for NGC~5806 \citep{cappellari2011}, whereby $1\arcsec$ corresponds to $\approx103$~pc.

\begin{figure*}
  \centering
  \includegraphics[width=2\columnwidth]{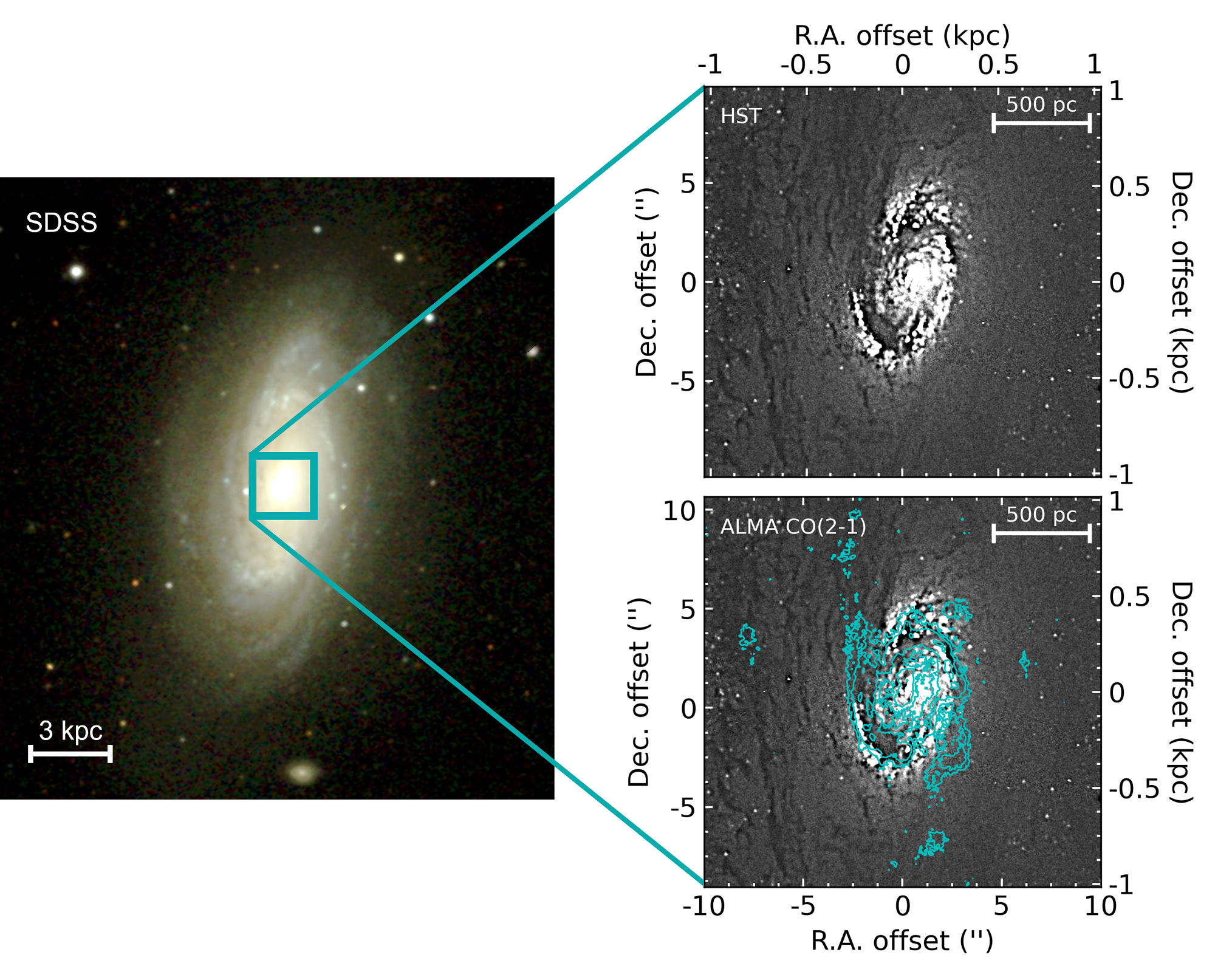}
  \caption{Left: SDSS three-colour ($gri$) image of NGC~5806, $2\farcm6\times2\farcm6$ ($16.4\times16.4$~kpc$^2$). Top-right: unsharp-masked \emph{HST} WFPC3 F555W image of a $2\times2$~kpc$^2$ region around the nucleus. Bottom-right: as above, but overlaid with cyan \CO\ integrated intensity contours from our ALMA observations. The molecular gas is co-spatial with the bright nucleus, nuclear ring and offset dust lanes.}
  \label{fig:optical_image}
\end{figure*}

NGC~5806 has a total stellar mass of $3.89\times10^{10}$~\Msun\ \citep{salo2017,morales2018}, a luminosity-weighted stellar velocity dispersion $\sigma_*=120$~\kms\ within the central $10\arcsec$ \citep{dumas2007}, an inclination $i=58\degr$ and a position angle $PA=166\degr$. The mass of molecular gas in the centre of NGC~5806 ($27\farcs4$ diameter) is $\approx10^9$~\Msun\ \citep{davis2022_wisdom} and the total mass of atomic hydrogen  \citep{haynes2018ALFALFA}. The \ion{H}{i} distribution traces the optical disc well \citep{mundell2007}. \autoref{fig:optical_image} shows the Sloan Digital Sky Survey (SDSS) three-colour image of NGC~5806 (left), a \textit{Hubble Space Telescope} (\textit{HST}) Wide-Field and Planetary Camera~3 (WFPC3) F555W image (top-right) and the \CO\ integrated intensity contours derived in \autoref{sec:data} overlaid on the same \textit{HST} image (bottom-right). On large scales, NGC~5806 has a large-scale bar, inner star-forming ring encirling the bar and weak spiral arms protruding from the bar. In the central region (i.e.\ well within the bar), NGC~5806 has a bright core and a star-forming nuclear ring that are prominent in both optical continuum and molecular gas emission. NGC~5806 has been classified as a Seyfert~2 galaxy \citep{dumas2007}, while more recent integral-field spectroscopic observations reveal ionised gas with mixed ionisation mechanisms \citep{westoby2007,westoby2012,errozferrer2019}. Star formation is present only in the inner and nuclear rings, with a total star-formation rate (SFR) of $3.6$~\Msun~yr$^{-1}$ derived using a spectral energy distribution fitting code \citep{errozferrer2019}. \citet{dumas2007} estimated the mass of the central supermassive black hole to be $\approx1.2\times10^7$~\Msun\, using the $M_{\mathrm{BH}}$ -- $\sigma_\star$ relation of \citet{tremaine2002}.

\subsection{Data}
\label{sec:data}

NGC~5806 was observed in the \CO\ line (rest frequency $230.586$~GHz) using ALMA as part of the WISDOM project. The observations were carried out using two different $12$-m array configurations in October and December 2016 (programme 2016.1.00437.S, configurations C40-3 and C40-6, PI Davis) and the $7$-m Atacama Compact Array (ACA) in July 2017 (programme 2016.2.00053.S, PI Liu) to achieve both high angular resolution and good flux recovery. The C40-3 configuration observations had $242$~s on-source using $44$ antennae and baselines of $15$ -- $600$~m, leading to a maximum recoverable scale of $6\farcs0$. The C40-6 configuration observations had $272$~s on-source using $41$ antennae and baselines of $15$ -- $1800$~m, leading to a maximum recoverable scale of $1\farcs3$. Both configurations have a primary beam of $27\farcs4$ (full-width at half-maximum; FWHM). The correlator was set up with one spectral window of $1.875$~GHz bandwidth ($\approx2400$~\kms) and $3840$ channels each of $488$~kHz ($\approx0.6$~\kms) used for the \CO\ line observations, and the three remaining spectral windows of $2$~GHz bandwidth used solely for continuum observations. The ACA observations had $1088$~s on-source using $10$ antennae and baselines of $8$ -- $43$~m, leading to a maximum recoverable scale of $29\farcs0$. The ACA observations have a primary beam of $45\farcs7$. The correlator was set up with one spectral window of $2$~GHz bandwidth ($\approx2600$~\kms) and $2048$ channels each of $977$~kHz ($\approx1.3$~\kms) used for the \CO\ line observations, and the three remaining spectral windows of $2$~GHz bandwidth used solely for continuum observations.

\subsubsection{Data reduction}

The raw data of each configuration were calibrated using the standard ALMA pipeline provided by ALMA regional centre staff, using \textsc{Common Astronomy Software Applications} (\textsc{casa}; \citealt{mcmullin2007_casa}) version 4.7.0. To combine the different configurations and obtain optimal sensitivity and spatial resolution for our science goals, we manually applied a low weighting ($0.2$) to the shorter baseline $12$-m data (C40-3) and a higher weighting ($1.0$) to the longer baseline $12$-m data (C40-6). Using \textsc{casa} version 5.6.1, we then combined the ACA data using the \textsc{casa} task \texttt{concat} with default weighting. Although continuum emission is not detected (see below), we subtracted any continuum that may be present using the \textsc{casa} task \texttt{uvcontsub}. We then cleaned the data using the \texttt{tclean} task interactively, to a depth equal to the root-mean-square (RMS) noise of the dirty cube, and imaged the cleaned components using Briggs weighting with a robust parameter of $0.5$. Finally, we achieved a synthesised beam of $\theta_{\mathrm{maj}}\times\theta_{\mathrm{min}}=$ $0\farcs25\times0\farcs22$ ($25.7\times22.6$~pc$^2$) at a position angle of $48\degr$. Pixels of $0\farcs05$ were chosen as a compromise between spatial sampling and image size, resulting in approximately $5\times4.5$ pixels across the synthesised beam. We thus created a fully calibrated and cleaned cube encompassing most of the primary beam spatially, with $2$~\kms\ (binned) channels spectrally. The RMS noise of this cube is $\sigma_{\mathrm{rms}}=0.86$~mJy~beam$^{-1}$ ($0.85$~K) per channel.

As mentioned above, no continuum emission is detected in NGC~5806. To establish an upper limit, we created a continuum image using the \texttt{tclean} task in \textsc{casa} and Briggs weighting with a robust parameter of $0.5$, resulting in a synthesised beam of $0\farcs20\times0\farcs18$. Averaging over the entire line-free bandwidth ($\approx6.3$~GHz), the resulting RMS noise is $25$~$\mu$Jy~beam$^{-1}$ at a central frequency of $238.351$~GHz.

\subsubsection{Moment maps}
 
\begin{figure*}
  \centering
  \includegraphics[height=0.90\columnwidth]{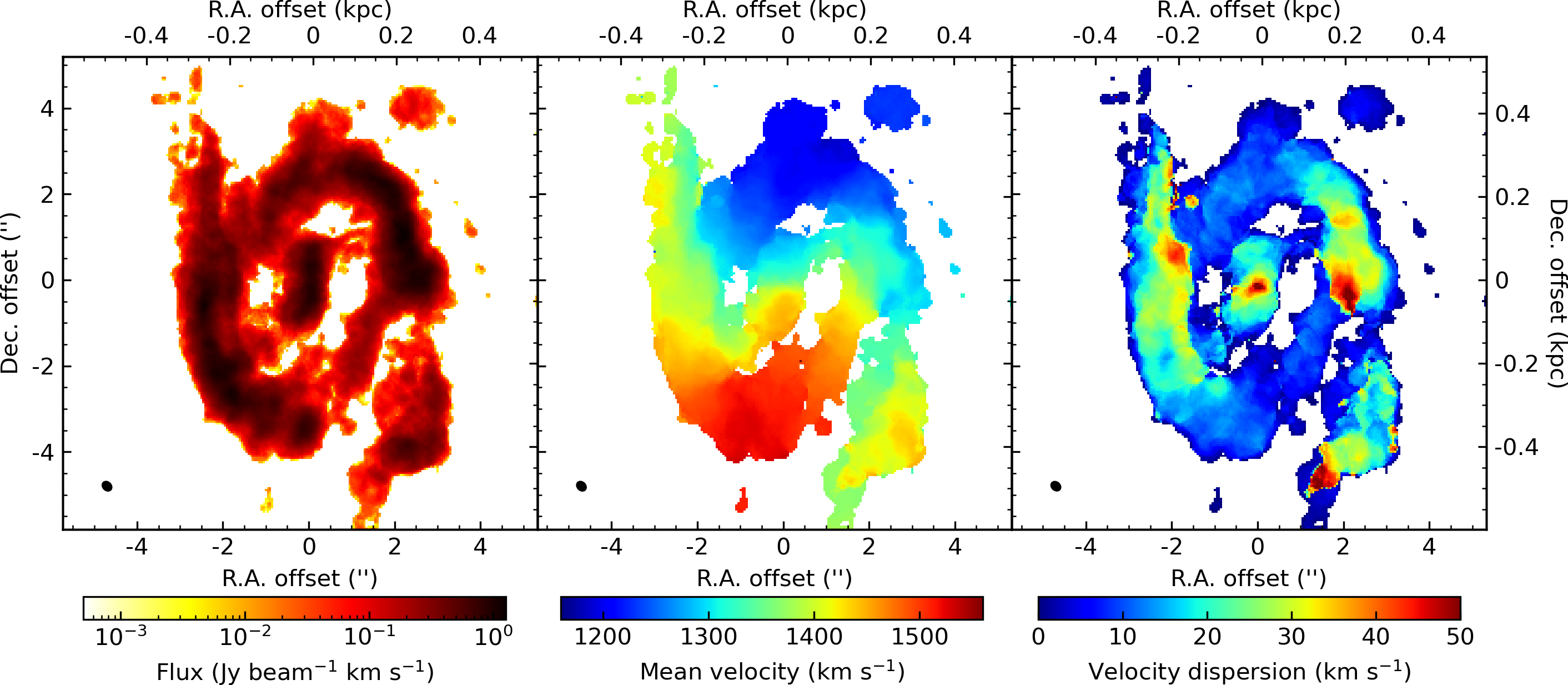}\hfill
  \includegraphics[height=0.77\columnwidth]{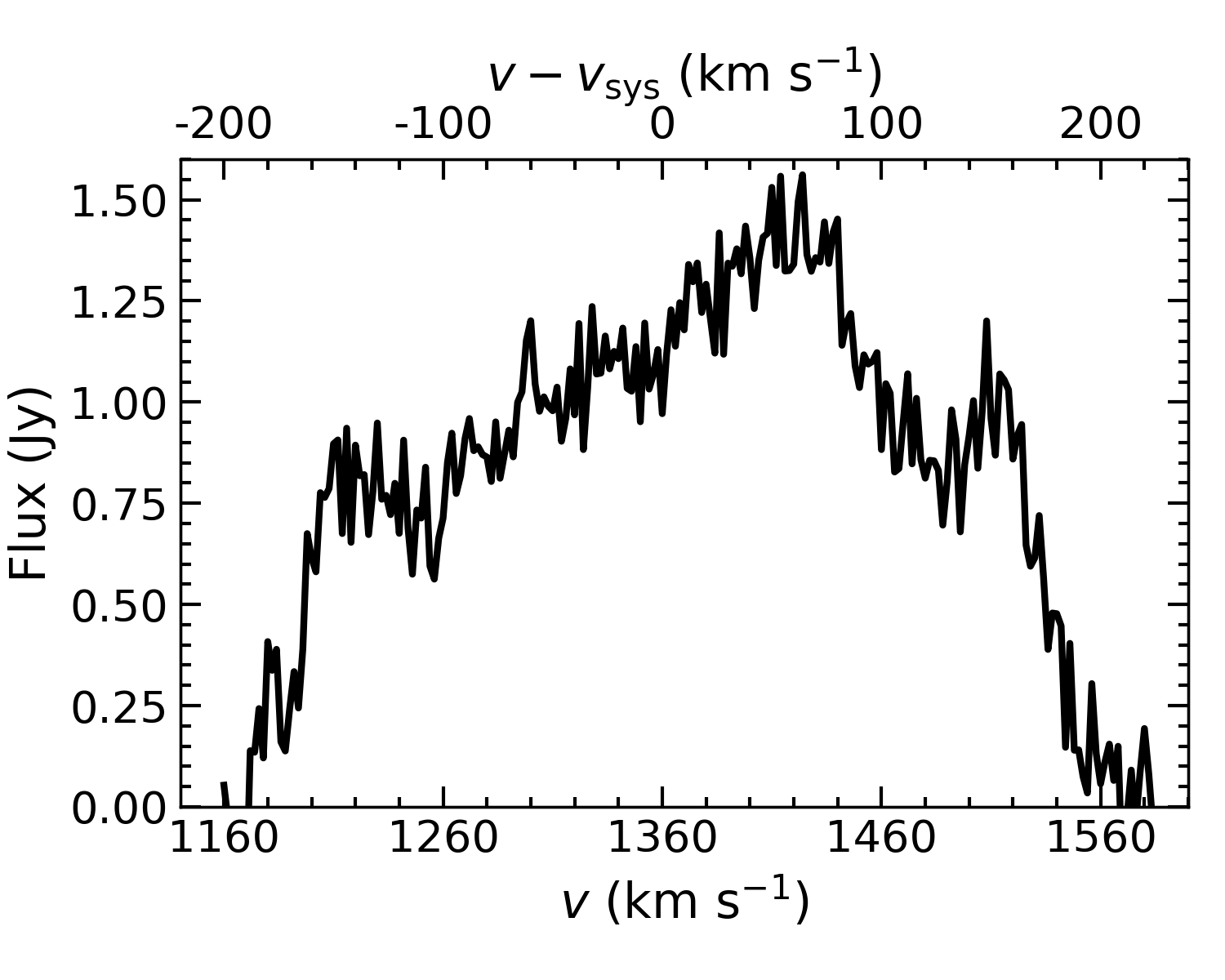}\hfill
  \caption{Moment maps of the \CO\ emission of NGC~5806. Top-left: zeroth-moment (integrated intensity) map. Top-middle: first-moment (intensity-weighted mean velocity) map. Top-right: second-moment (intensity-weighted velocity dispersion) map. Bottom: Integrated \CO\ spectrum, extracted from a $9\arcsec\times9\arcsec$ region around the galaxy centre. The synthesised beam of $0\farcs25\times0\farcs22$ ($25.7\times22.6$~pc$^2$) is shown in the bottom-left corner of each moment map.}
  \label{fig:moments}
\end{figure*}

\autoref{fig:moments} shows the zeroth-moment (total intensity) map (top-left), first-moment (intensity-weighted mean velocity) map (top-middle) and second-moment (intensity-weighted velocity dispersion) map (top right) of the \CO\ line of NGC~5806. To generate these maps, we utilised a smooth-moment masking method \citep[e.g.][]{dame2011}. In brief, we convolved the data cube spatially with a Gaussian of width equal to that of the synthesised beam and Hanning-smoothed the cube spectrally. We then only selected pixels with an intensity above $1.5$ times the RMS noice of the smoothed cube to create a mask, and applied this mask to the original data cube to create the moment maps.

The integrated intensity map reveals a highly structured molecular gas distribution. In particular, molecular gas is associated with the nucleus at the very centre of the galaxy, the particularly dusty part of the bright optical nuclear ring and the bi-symmetric offset dust lanes of the large-scale bar (stretching to the north and south; see \autoref{fig:optical_image}). In addition, the integrated intensity is high at the interfaces between the offset dust lanes and the nuclear ring, and it decreases gradually as a function of the azimuthal angle in a counter-clockwise direction.

The mean velocity map clearly shows that the northern side of the ring is blue-shifted while the southern side is red-shifted with respect to the systemic velocity $V_{\mathrm{sys}}=1360$~\kms (as determined from \ion{H}{i} line emission; \citealt{springob2005_vsys_ref}). The eastern and western sides of the ring also show blue- and red-shifted velocities along the spiral arms, indicating deviations from circular motions, leading to a complex velocity field.

The velocity dispersion of the molecular gas is generally higher ($0$ -- $60$~\kms) than that of nearby galaxies \citep[e.g.][]{wilson2011_vel_disp,mogotsi2016_vel_disp,sun2018_vel_disp}. In particular, the velocity dispersions at the interfaces between the offset dust lanes and the nuclear ring are higher ($30$ -- $50$~\kms) that those in other parts of the nuclear ring ($0$ -- $20$~\kms), indicating that these environments are likely to be different from each other. The nucleus also shows high velocity dispersions ($30$ -- $60$~\kms). The complex velocity field and high velocity dispersions of NGC~5806 are further discussed in \autoref{sec:discussion_turbulence}.

The bottom-left panel of \autoref{fig:moments} shows the integrated CO spectrum of a $9\arcsec\times9\arcsec$ central region, revealing multiple peaks and thus suggesting again complex molecular gas distribution and kinematics. The total \CO\ flux in that region is $\approx300$~Jy~\kms.

\subsubsection{Region definitions}
\label{sec:regions}

\begin{figure*}
  \centering
  \includegraphics[width=1.4\columnwidth]{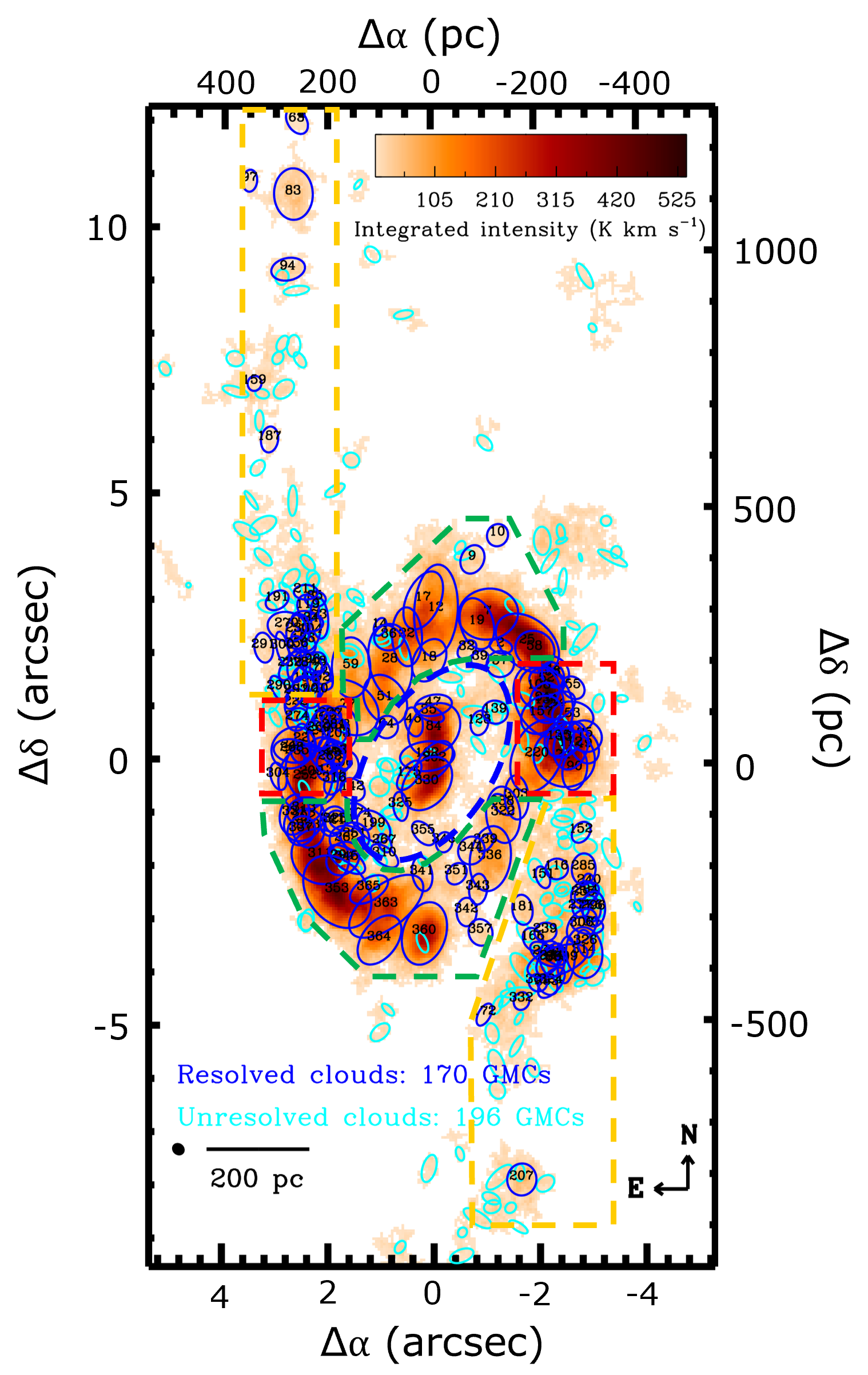}
  \caption{\CO\ integrated intensity map of NGC~5806 with identified GMCs overlaid. Dark blue (cyan) ellipses indicate resolved (unresolved) clouds. Blue (nucleus), green (arcs), red (nodes) and yellow (dust lanes) polygons indicate the four regions defined in \autoref{sec:regions}.}
  \label{fig:cloudfinding_result}
\end{figure*}

Based on the moment maps, we divide the galaxy into four distinct regions, referred to as follows (see \autoref{fig:cloudfinding_result}): nucleus (blue), arcs (green), nodes (red) and dust lanes (yellow). The nucleus encompasses only the inner $125$~pc in radius, the arcs refer to the parts of the nuclear ring where the velocity dispersions are relatively low, the nodes refer to the parts of the nuclear ring that are at the interfaces between the nuclear ring and the offset dust lanes and where the velocity dispersions are relatively high, and the dust lanes indicate the offset dust lanes in the optical image that are characteristic of barred disc galaxies \citep[e.g.][]{anthanassoula1992_dust_shock}. We note that we will refer to the nuclear ring only to refer to the nuclear ring in its entirety, encompassing both the arcs and the nodes.

\subsection{Cloud identification}
\label{sec:identification}

We utilise our own modified version of the algorithms of \textsc{cpropstoo} \citep{liu2021ngc4429}, that is an updated version of \textsc{CPROPS} \citep{Rosolowsky2006cprops,Leroy2015cpropstoo}, to identify the clouds of NGC~5806. Our version of \textsc{cpropstoo} has fewer free parameters, leading to a more efficient and robust cloud identification in complex and crowded environments. We refer the reader to \citet{liu2021ngc4429} for full details of our version of \textsc{cpropstoo}.

We introduce here the main steps and parameters of the algorithm. First, the algorithm calculates the spatially-varying noise in the cube and generates a three-dimensional (3D) mask of bright emission. The mask initially includes only pixels for which two adjacent channels are above $2.5$~\rms. The mask is then expanded to include all neighbouring pixels for which two adjacent channels are above $1.5$~\rms. The individual regions identified are referred to as "islands". To remove noise peaks, we exclude all islands with projected areas less than two synthesised beams. We also apply the same criteria to the inverted data cube to verify the reliability of our island identification.

Second, the islands identified are decomposed into individual structures, that we refer to as clouds. Local maxima (i.e.\ cloud candidates) are identified within running $3\times3\times3$~pix$^3$ subsets of the cube (i.e.\ $0\farcs15\times0\farcs15\times6$~\kms\ sub-cubes). To eliminate noise peaks and outliers, we also require the total emission in each $3\times3\times3$~pix$^3$ sub-cube to be greater than that in the eight spatially-neighbouring sub-cubes. We then run \textsc{cpropstoo}, setting the minimum number of channels spanned by each cloud ($minvchan=2$) and the minimum contrast between a cloud's peak and its boundary ($\Delta T_{\mathrm{max}}=2\,\sigma_{\mathrm{rms}}=1.7$~K).

Individual cloud candidates have to occupy a minimum area within which all emission is uniquely associated as dictated by two parameters: $minarea$ (minimum cloud area) and $minpix$ (minimum number of pixels). However, biases can occur depending on $minarea$ and $minpix$, e.g.\ small structures may be missed when these two parameters are set high, whereas large structures may be missed when they are set low. To minimise this potential bias, rather than using a single value we assign both parameters a range of $96$ -- $24$ spaxels (i.e.\ the synthesised beam area). The code searches for clouds from the largest $minarea$ ($96$ spaxels) and $minpix$ ($96$ pixels) to the smallest $minarea$ ($24$ spaxels) and $minpix$ ($24$ pixels) with a step size of $24$ spaxels (or pixels). This modification allows to reduce the arbitrariness of the search area.

To counteract the weakness of the algorithm, that is likely to ignore significant sub-structures of large clouds, \citet{liu2021ngc4429} introduced an additional parameter, $convexity$, defined as the ratio of the volume of a cloud's 3D intensity distribution to that of its convex envelope. When $convexity\approx1$, the cloud has only one intensity peak, while the smaller the $convexity$ the more significant the sub-structures. In this work, we set $convexity=0.5$ by testing a range of $0.4$ -- $0.8$. Values in the range $0.5$ -- $0.7$ are typical \citep{liu2021ngc4429}. This parameter allows to identify structures over multiple scales with less arbitrariness.

As a result, we identify $366$ GMCs, $170$ of which are both spatially and spectrally resolved, as shown in \autoref{fig:cloudfinding_result}. We note that two resolved clouds do not belong to any of the four regions defined.

\section{Cloud properties}
\label{sec:properties}

\subsection{GMC properties}

Following the standard \textsc{cpropstoo}/\textsc{cprops} definitions \citep{Rosolowsky2006cprops}, we calculate the physical properties of the clouds identified. We list the (intensity-weighted) properties of each cloud in \autoref{tab:table1}, including each cloud's central position (R.A.\ and Dec.), mean local standard of rest velocity ($V_{\mathrm{LSR}}$), size (radius \Rc), observed velocity dispersion (\sigobs), gradient-subtracted velocity dispersion (\siggs; see \citealt{liu2021ngc4429}), \CO\ luminosity (\Lco), molecular gas mass (\Mgas), peak intensity ($T_{\mathrm{max}}$), projected angular velocity (\omeobs), position angle of the rotation axis (\rotaxis; see \autoref{sec:gradients}) and deprojected distance from the galaxy centre ($R_{\mathrm{gal}}$). Some quantities are discussed below, but see also \citet{liu2021ngc4429}.

\begin{table*}
  \centering
  \caption{Observed properties of the clouds of NGC~5806. A complete machine-readable version of this table is available in the online journal version.}
  \label{tab:table1}
  \resizebox{\textwidth}{!}{
    \begin{tabular}{ccccccccccccc}
      \hline
      ID & RA (2000) & Dec.\ (2000) & $V_{\mathrm{LSR}}$  & $R_{\mathrm{c}}$ & $\sigma_{\mathrm{obs,los}}$ & $\sigma_{\mathrm{gs,los}}$ & $L_{\mathrm{CO(2-1)}}$ & $M_{\mathrm{gas}}$ & $T_{\mathrm{max}}$ & $\omega_{\mathrm{obs}}$  & $\phi_{\mathrm{rot}}$ & $R_{\mathrm{gal}}$ \\
         & (h:m:s) & ($^{\circ}:^\prime:^{\prime\prime}$) & (\kms) & (pc) & (\kms) & (\kms) & ($10^4$~K~km~s$^{-1}$~pc$^{-2}$) & ($10^5$~M$_\odot$) & (K) & (\kms~pc$^{2}$) & ($^{\circ}$) & (pc) \\
      \hline
      1   & 15:00:0.31  & 1:53:31.58  & 1168.9     & -     & 1.37 $\pm$ 3.14       & -       & 0.97 $\pm$ 1.87                       & 0.43 $\pm$ 0.82                        & 3.2     & -          & -         & 336      \\
      2   & 15:00:0.33  & 1:53:30.83  & 1180.5     & 16.06 $\pm$ 24.54    & 2.23 $\pm$ 1.78       & 2.12 $\pm$ 3.06      & 2.70 $\pm$ 1.18   & 1.19 $\pm$ 0.52  & 4.2     & 0.06 $\pm$ 0.05 & 256 $\pm$ 121       & 254      \\
      3   & 15:00:0.50  & 1:53:31.67  & 1184.0     & -     & -       & -       & 0.38   $\pm$ 1.04                     & 0.17 $\pm$ 0.46                       & 3.2     & -          & -         & 522      \\
      4   & 15:00:0.37  & 1:53:32.55  & 1192.4     & -     & 2.30 $\pm$ 2.38       & 1.61 $\pm$ 2.03       & 1.16 $\pm$ 0.74                        & 0.51 $\pm$ 0.33                        & 3.3     & -          & -         & 412      \\
      5   & 15:00:0.20  & 1:53:33.19  & 1194.0     & -     & -       & -       & 0.72 $\pm$ 0.93                       & 0.32 $\pm$ 0.41                        & 3.8     & -          & -        & 633      \\
      6   & 15:00:0.48  & 1:53:30.94  & 1203.4     & -     & 3.56 $\pm$ 1.40      & 2.12 $\pm$ 1.73       & 1.98 $\pm$ 0.54                       & 0.87 $\pm$ 0.24                       & 4.5     & -          & -        & 393      \\
      7   & 15:00:0.34  & 1:53:31.44  & 1203.3     & 62.57 $\pm$ \phantom{0}1.83    & 15.12 $\pm$ 0.47\phantom{0}      & 13.0 $\pm$ 0.46     & 243.09  $\pm$ 5.14\phantom{0}\phantom{0} & 106.96 $\pm$ 2.26\phantom{0}\phantom{0} & 10.5\phantom{0}    & 0.19 $\pm$ 0.01 & 260 $\pm$ \phantom{0}\phantom{0}1       & 293      \\
      8   & 15:00:0.25  & 1:53:32.82  & 1206.3     & -     & 2.03 $\pm$ 2.57       & 1.23 $\pm$ 1.84       & 0.99 $\pm$ 1.37                       & 0.44 $\pm$ 0.60                      & 3.6     & -          & -        & 513      \\
      9   & 15:00:0.36  & 1:53:32.47  & 1207.2     & 25.18 $\pm$ 11.30   & 4.81 $\pm$ 1.75      & 3.32 $\pm$ 2.40      & 13.35  $\pm$ 4.30\phantom{0}   & 5.87 $\pm$ 1.89  & 4.8     & 0.10 $\pm$ 0.05 & 296 $\pm$ \phantom{0}17       & 399      \\
      10  & 15:00:0.33  & 1:53:32.92  & 1210.8     & 17.59 $\pm$ \phantom{0}7.67    & 7.39 $\pm$ 3.55      & 3.73 $\pm$ 2.66       & 4.32   $\pm$ 0.89  & 1.90 $\pm$ 0.39  & 4.7     & 0.36 $\pm$ 0.08 & 296 $\pm$ \phantom{0}41       & 447      \\
      11  & 15:00:0.47  & 1:53:31.00  & 1211.5     & -     & 2.53 $\pm$ 2.05      & 1.11 $\pm$ 3.17      & 1.26 $\pm$ 3.28                       & 0.55 $\pm$ 1.44                       & 4.4     & -          & -       & 393      \\
      12  & 15:00:0.41  & 1:53:31.50  & 1206.5     & 74.01 $\pm$ \phantom{0}2.28    & 9.69 $\pm$ 0.36      & 8.83 $\pm$ 0.40      & 194.23  $\pm$ 4.42\phantom{0}\phantom{0}  & 85.46 $\pm$ 1.95\phantom{0} & 10.4\phantom{0}    & 0.09 $\pm$ 0.01 & 196 $\pm$ \phantom{0}\phantom{0}1       & 325      \\
      13  & 15:00:0.38  & 1:53:32.69  & 1211.0     & -     & 1.43 $\pm$ 2.29       & 0.83 $\pm$ 2.24      & 2.52 $\pm$ 2.03                       & 1.11 $\pm$ 0.89                       & 4.9     & -          & -        & 430      \\
      14  & 15:00:0.48  & 1:53:31.20  & 1214.2     & 19.24 $\pm$ 19.66    & 2.93 $\pm$ 2.32      & 1.79 $\pm$ 3.46      & 3.84 $\pm$ 3.39    & 1.69 $\pm$ 1.49  & 4.0     & 0.09 $\pm$ 0.11 & 243 $\pm$ 128       & 419      \\
      15  & 15:00:0.24  & 1:53:31.22  & 1211.3     & -     & 4.21 $\pm$ 5.96      & -       & 2.63 $\pm$ 6.92                       & 1.16 $\pm$ 3.05                       & 3.3     & -           & -         & 453      \\
      
      -&-&-&-&-&-&-&-&-&-&-&-&\\
      366 & 15:00:0.43  & 1:53:25.27  & 1554.1     & -     & 1.74 $\pm$ 2.26      & 0.88 $\pm$ 1.76      & 1.00 $\pm$ 1.10                        & 0.44 $\pm$ 0.48                       & 4.1     & -          & -         & 366 \\
      \hline
    \end{tabular}
  }
\end{table*}

The cloud size (\Rc) is defined as
\begin{equation}
  R_{\mathrm{c}}\equiv\eta\sqrt{\sigma_{\mathrm{maj,dc}}\,\sigma_{\mathrm{min,dc}}}\,\,\,,
\end{equation}
where $\eta$ is a geometric parameter, $\sigma_{\mathrm{maj,dc}}$ and $\sigma_{\mathrm{min,dc}}$ are the deconvolved RMS spatial extent along the major and the minor axis of the cloud, respectively, and we adopt $\eta=1.91$ for consistency with earlier studies \citep[e.g.][]{solomon1987,utomo2015,liu2021ngc4429}.

The observed velocity dispersion (\sigobs) is calculated as
\begin{equation}
  \sigma_{\mathrm{obs,los}}\equiv\sqrt{\left(\sigma_{\mathrm{v}}^2-(\Delta V_{\mathrm{chan}}^2/2\pi)\right)}\,\,\,,
\end{equation}
where $\sigma_{\mathrm{v}}$ is the second velocity moment and $\Delta V_{\mathrm{chan}}$ the channel width of the data cube.

The molecular gas mass (\Mgas) is calculated from the \CO\ luminosity (\Lco), itself obtained from the \CO\ flux (\Fco) by
\begin{equation}
  \left(\frac{L_{\mathrm{CO(2-1)}}}{\mathrm{K~km~s^{-1}~pc^2}}\right)=\left(\frac{3.25\times10^7}{(1+z)^3}\right)\left(\frac{F_{\mathrm{CO(2-1)}}}{\mathrm{Jy~km~s^{-1}}}\right)\left(\frac{\nu_{\mathrm{obs}}}{\mathrm{GHz}}\right)^{-2}\left(\frac{D}{\mathrm{Mpc}}\right)^2\,\,\,,
\end{equation}
where $z$ is the galaxy redshift and $\nu_{\mathrm{obs}}$ the observed line frequency. To convert the CO luminosity to a molecular gas mass, we adopt a $^{12}$CO(2-1)/$^{12}$CO(1-0) ratio $R_{21}=1$ in temperature units, within the range typically found in the central regions of (barred) spiral galaxies ($0.8$ -- $1.2$; e.g.\ \citealt{Crosthawaite2002_lineratio}), and a CO-to-H2 conversion factor $X_{\mathrm{CO}}=2\times10^{20}$~cm$^{-2}$~(K~\kms)$^{-1}$ equivalent to a \CO\ conversion factor $\alpha_{\mathrm{CO(2-1)}}\approx4.4$~M$_{\odot}$~(K~km~s$^{-1}$~pc$^2$)$^{-1}$. This yields
\begin{equation}
  \begin{split}
    \left(\frac{M_{\mathrm{gas}}}{\mathrm{M_\odot}}\right) & = \frac{X_{\mathrm{CO}}}{R_{21}}\left(\frac{L_{\mathrm{CO(2-1)}}}{\mathrm{K~km~s^{-1}~pc^2}}\right)\\
    & \approx\ 4.4\,\left(\frac{L_{\mathrm{CO(2-1)}}}{\mathrm{K~km~s^{-1}~pc^2}}\right)\,\,\,.\\
  \end{split} 
\end{equation}

In this work, we also use a second measure of the velocity dispersion, the gradient-subtracted velocity dispersion \siggs\ introduced in previous GMC studies \citep{utomo2015,liu2021ngc4429}. This quantity is calculated as follows. First, we calculate the intensity-weighted mean velocity at each spaxel of a cloud, and measure its offset with regards to the mean velocity at the cloud centre. Second, we shift the spectrum at each spaxel to match its mean velocity to that of the cloud centre. Finally, we calculate the second moment of the shifted emission summed over the whole cloud, and extrapolate it to $T_{\mathrm{edge}}=0$~K. This new \siggs\ measure thus quantifies the turbulent motions within the cloud, with any bulk motion removed.

The uncertainties of all cloud properties are estimated via a bootstrapping technique as in \citet{liu2021ngc4429}, with $500$ samples. The uncertainty of the galaxy distance $D$ is not considered, as an error of the distance translates to a systematic scaling of some quantities, i.e.\ $R_{\mathrm{c}}\propto D$, $L_{\mathrm{CO(2-1)}}\propto D^2$, $M_{\mathrm{gas}}\propto D^2$, $\omega_{\mathrm{obs}}\propto D^{-1}$, $M_{\mathrm{vir}}\propto D$ (\autoref{sec:dynamics}) and $R_{\mathrm{gal}}\propto D$.

\subsection{Distributions of GMC properties}

\begin{figure*}
  \centering
  \includegraphics[width=2\columnwidth]{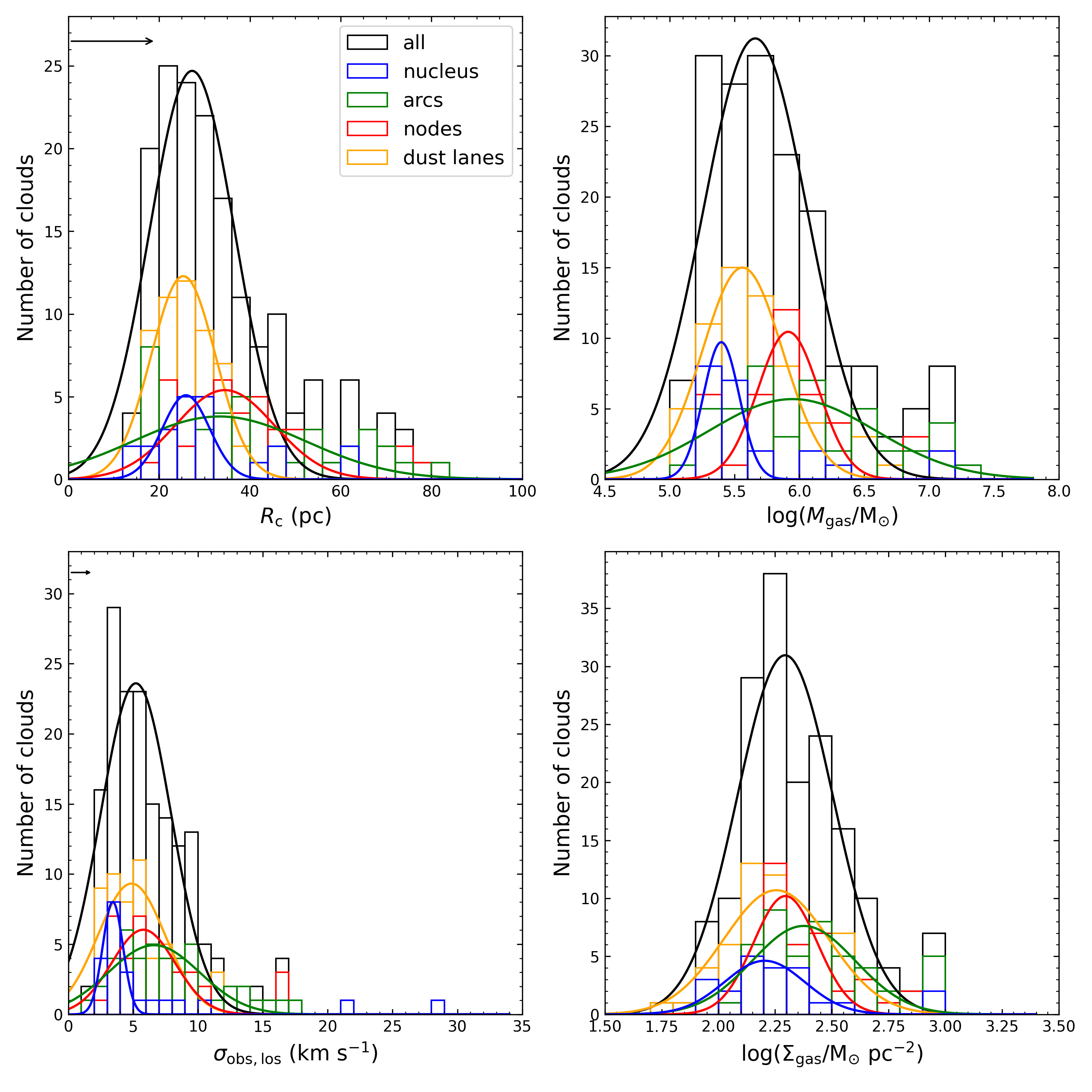}
  \caption{Number distributions of \Rc, $\log(M_{\mathrm{gas}}/\mathrm{M_{\odot}})$, \sigobs~and $\log(\Sigma_{\mathrm{gas}}/\mathrm{M_\odot~pc^{-2}})$ with their Gaussian fits overlaid for the $170$ resolved clouds of NGC~5806 (black lines and histograms), and for the clouds in the nucleus (blue), arcs (green), nodes (red) and dust lanes (yellow) only. The black arrows in the top-left and bottom-left panels indicate our ability to resolve clouds spatially ($\eta\sqrt{\sigma_{\mathrm{maj}}\,\sigma_{\mathrm{min}}}$, where $\sigma_{\mathrm{maj,min}}\equiv\theta_{\mathrm{maj,min}}/2.35$) and spectrally (channel width of $2$~km~s$^{-1}$), respectively.}
  \label{fig:histogram}
\end{figure*}

\autoref{fig:histogram} shows the number distributions of cloud size (\Rc), gas mass (\Mgas), observed velocity dispersion (\sigobs) and gas mass surface density ($\Sigma_{\mathrm{gas}}\equiv M_{\mathrm{gas}}/\pi R_{\mathrm{c}}^2$) for the resolved clouds of NGC~5806. As described above, we divide the clouds into four groups, one for each spatial region within the galaxy. In each panel, the black histogram (data) and curve (Gaussian fit) show the full sample, while the blue, green, red and yellow colours show those of the clouds in the nucleus, arcs, nodes and offset dust lanes only, respectively.

The sizes (\Rc) of the resolved clouds of NGC~5806 range from $15$ to $85$~pc (top-left panel of \autoref{fig:histogram}). The mean of the Gaussian fit is $27.8\pm0.7$~pc and the standard deviation $9.4$~pc, while the median radius is $30.2$~pc. The resolved clouds have gas masses \Mgas\ ranging from $1.2\times10^5$ to $3.6\times10^7$~\Msun\ (top-right panel of \autoref{fig:histogram}). The mean of the Gaussian fit to the $\log(M_{\mathrm{gas}}/\mathrm{M_{\odot}})$ distribution is $5.66\pm0.04$ ($\approx4.6\times10^5$~\Msun) and the standard deviation $0.4$, while the median gas mass is $5.5\times10^5$~\Msun. About one third ($49/170$) of the resolved clouds are massive ($M_{\mathrm{gas}}\geq10^6$~\Msun). The observed velocity dispersions of the resolved clouds range from $1.6$ to $30$~\kms\ (bottom-left panel of \autoref{fig:histogram}). The mean of the Gaussian fit is $5.2\pm0.2$~\kms\ and the standard deviation $2.7$~\kms, while the median observed velocity dispersion is $5.6$~\kms. The gas mass surface densities of the resolved clouds range from $80$ to $1000$~\Msun~pc$^{-2}$ (bottom-right panel of \autoref{fig:histogram}). The mean of the Gaussian fit to the $\log(\Sigma_{\mathrm{gas}}/{\mathrm{M_\odot~pc^{-2}}})$ distribution is $2.29\pm0.02$ ($\approx195$~\Msun) and the standard deviation $0.2$, while the median gas mass surface density is $2.3$ ($\approx200$~\Msun).

There are slight variations of all four quantities across the four regions. The clouds in the arcs and nodes tend to be larger than the clouds in the nucleus and offset dust lanes (median radius $\approx38$ and $36$~pc vs.\ $\approx27$ and $27$~pc), more massive (median gas mass $\approx10^{6.1}$ and $10^{5.9}$~\Msun\ vs.\ $\approx10^{5.5}$ and $10^{5.6}$~\Msun) and more turbulent (median observed velocity dispersion $\approx7.5$ and $6.1$~\kms\ vs.\ $\approx3.6$ and $5.1$~\kms). We also identified two clouds that have exceptionally large velocity dispersions ($\approx21$ and $29$~\kms) in the nucleus, despite not being the largest and/or most massive clouds, indicating that those clouds are likely to be affected by their surrounding environment, e.g.\ the active galactic nucleus (AGN) and/or strong galactic shear. The median gas mass surface density of the clouds in the arcs ($\langle\Sigma_{\mathrm{gas}}\rangle\approx280$~\Msun~pc$^{-2}$) is larger than that of the clouds in the other three regions ($\langle\Sigma_{\mathrm{gas}}\rangle\approx190$~\Msun~pc$^{-2}$).

The resolved clouds of NGC~5806 have sizes comparable to and masses slightly larger than those of the clouds in the MWd ($R_{\mathrm{c}}=30$ -- $50$~pc and $M_{\mathrm{gas}}=10^{4.5}$ -- $10^{7.5}$~\Msun, with $\leq20$~pc spatial resolution; \citealt{rice2016mw,miville2017a}), but they have sizes and masses larger than those of the clouds in the central molecular zone (CMZ; $R_{\mathrm{c}}=5$ -- $15$~pc and $M_{\mathrm{gas}}=10^{3.3}$ -- $10^{6}$~\Msun, with $\leq1.5$~pc resolution; \citealt{oka2001,kauffmann2017}). On the contrary, the velocity dispersions of the NGC~5806 clouds are slightly larger and smaller than those of the clouds in the MWd ($1$ -- $6$~\kms; \citealt{heyer2009}) and the CMZ ($12$ -- $50$~\kms; \citealt{oka1998}), respectively.  Most clouds in late-type galaxies have comparable sizes ($20$ -- $200$~pc), masses ($10^{4.5}$ -- $10^{7.5}$~\Msun) and observed velocity dispersions ($2$ -- $10$~\kms; $10$ -- $60$~pc resolution; e.g.\ \citealt{donovan2012,hughes2013,rebolledo2015,liu_2023}), while clouds in ETGs have slightly smaller sizes ($5$ -- $50$~pc) and masses ($10^{4.4}$ -- $10^{6.6}$~\Msun) but comparable observed velocity dispersions ($2$ -- $20$~\kms) to those of the clouds in NGC~5806 ($\leq20$~pc resolution; \citealt{utomo2015,liu2021ngc4429}).

Overall, the clouds in the nucleus generally are the smallest, least massive, least turbulent and have the smallest surface densities. On the other hand, the clouds in the arcs and nodes are the largest, most massive, most turbulent and have the largest surface densities. The clouds in the offset dust lanes have intermediate properties.

\subsection{GMC cumulative mass functions}
\label{sec:massspec}

The mass function of GMCs is a tool to diagnose GMC populations and provides constraints on GMC formation and destruction \citep[e.g.][]{rosolowsky2005,colombo2014m51}. Here we use the gas mass rather than the virial mass to calculate the mass function, as the former is well defined even for spatially-unresolved clouds, and no assumption on the dynamical state of the clouds is required.

The cumulative mass functions are fit with both a power-law function
\begin{equation}
  N(M'>M)=\left(\frac{M}{M_0}\right)^{\gamma+1}\,\,\,,
\end{equation}
where $N(M'>M)$ is the number of clouds with a mass greater than $M$, $M_0$ sets the normalisation and $\gamma$ is the power-law index, and a truncated power-law function
\begin{equation}
  N(M'>M)=N_0\left[\left(\frac{M}{M_0}\right)^{\gamma+1}-1\right]\,\,\,,
\end{equation}
where $M_0$ is now the cut-off mass and $N_0$ is the number of clouds with a mass $M>2^{1/(\gamma+1)}\,M_0$. To fit each cumulative mass function, we apply the "error in variable" method of \citet{rosolowsky2005}, and the fitting parameters and the uncertainties are estimated via bootstrapping. Fittings are only performed above the mass completeness limit of $M_{\mathrm{comp}}=2.4\times10^5$~\Msun. We calculate the mass completeness limit using the minimum mass ($M_{\mathrm{min}}$) of resolved cloud and the observational sensitivity, i.e.\ $M_{\mathrm{comp}}\equiv M_{\mathrm{min}}+10\delta_{\mathrm{M}}$ \citep[e.g.][]{colombo2014m51,liu2021ngc4429}, where the the contribution to the mass due to noise $\delta_{\mathrm{M}}=1.03\times10^4$~\Msun\ is estimated by multiplying our RMS gas mass surface density sensitivity of $17.8$~\Msun~pc$^{-2}$ by the synthesised beam area of $565$~pc$^2$.

\begin{figure}
  \includegraphics[width=\columnwidth]{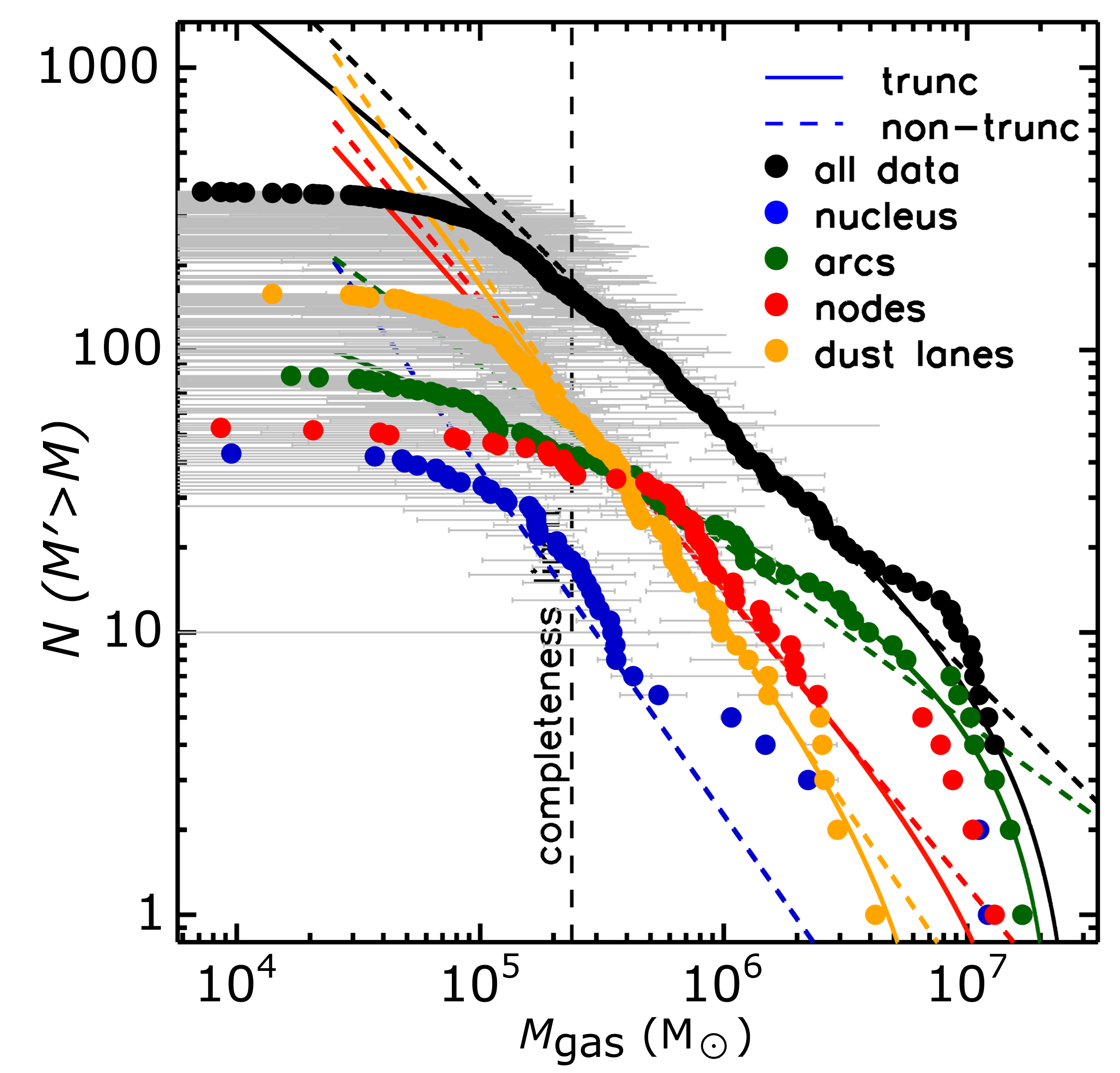}
  \caption{Cumulative gas mass distribution of all the clouds of NGC~5806 (black data points) and of the clouds in the nucleus (blue), arcs (green), nodes (red) and dust lanes (yellow) only. Truncated (solid lines) and non-truncated (dashed lines) power-law fits are overlaid. The mass completeness limit is shown as a black vertical dashed line.}
  \label{fig:mass_spectrum}
\end{figure}

\autoref{fig:mass_spectrum} shows the cumulative mass function of all identified clouds (black data points), with the best-fitting truncated power-law (black solid line) and non-truncated power-law (black dashed line) overlaid. The mass functions of the clouds in each regions are also shown in colours. The best-fitting slopes of the truncated and non-truncated power laws are $\gamma=-1.72\pm0.12$ and $\gamma=-1.86\pm0.06$, respectively. Although both the truncated and non-truncated power laws do not fit well at large masses due to the bump around $10^7$~\Msun, both slopes are shallower than that of the mass function of the clouds in the MWd ($-2.20\pm0.1$; \citealt{rice2016mw}), M~51 ($-2.30\pm1$; \citealt{colombo2014m51}), the outer regions of M~33 ($-2.10\pm1$; \citealt{rosolowsky2007etal}) and the ETGs NGC~4526 ($2.39\pm0.03$; \citealt{utomo2015}) and NGC~4429 ($-2.18\pm 0.21$; \citealt{liu2021ngc4429}), but they are similar to those of the clouds in the MW centre ($-1.60\pm0.1$; \citealt{rice2016mw}), the spiral arms of M~51 ($-1.79\pm0.09$; \citealt{colombo2014m51}), the inner regions of M~33 ($-1.80\pm1$; \citealt{rosolowsky2007etal}), NGC~300 ($-1.80\pm0.07$; \citealt{faesi2016}) and Local Group galaxies ($\approx-1.7$; \citealt{blitz2007}).

The GMC mass functions of the different regions are somewhat different from each other. We note that to avoid a failure of some of the fits, the five most massive clouds from each of the nucleus and the nodes were excluded from the fits (although the nucleus truncated power-law fit still fails). The GMCs of the nucleus, arcs, nodes and dust lanes have a best-fitting non-truncated power-law slope of $\gamma$ of $-2.22\pm0.39$, $-1.63\pm0.06$, $-2.04\pm0.27$ and $-2.27\pm0.14$, respectively. Due to limited number of clouds (i.e.\ the small sample size) in each region, and the fact that our modified \textsc{cpropstoo} code identifies GMCs with a fixed convexity over multiple scales (leading to bumpy mass functions), our best-fitting slopes for each region have large uncertainties and the fits do not always seem to represent the mass functions well, especially in the nucleus and nodes. Despite these limitations, however, the cloud mass functions of the nodes and arcs (i.e.\ nuclear ring) are significantly shallower than those of the nucleus and dust lanes. The former have a slope shallower than or close to $-2$, while the latter have a slope steeper than $-2$. This implies that massive GMCs preferentially reside along the nuclear ring, whereas the mass budgets in the nucleus and dust lanes are dominated by less-massive GMCs. It also suggests either the dust lanes lack an efficient cloud growth mechanism or they have an efficient cloud destruction mechanism.

It seems that the evolution and formation of GMCs are influenced by different galactic environments, and thus different GMC populations may exist in the galaxy. In M~51, \citet{colombo2014m51} also reported that the galactic environment can affect not only the physical properties of the clouds but also their cumulative mass function, reporting a sharp truncation of the mass function at high masses ($\approx10^{6.5}$~M$_\odot$) in the nuclear bar ($\approx1$~kpc diameter) compared to other regions (e.g.\ spiral arms). They suggested that galactic shear is likely to be a main driver of cloud destruction in the nuclear bar. In any case, both these results and ours imply that the galactic environment can influence the evolution and formation of GMCs.

\section{Cloud kinematics}
\label{sec:kinematics}

\subsection{Velocity gradients of individual clouds}
\label{sec:gradients}

Previous GMC studies have shown that the velocity gradients of GMCs can reveal internal cloud rotation \citep[e.g.][]{blitz1993,phillips1999,rosolowsky2003,rosolowsky2007,utomo2015,liu2021ngc4429}. Because clouds in external galaxies are usually poorly spatially resolved, solid-body rotation provides an adequate description of the observations. As previous studies, we thus quantify the observed velocity gradient by fitting a plane to the intensity-weighted first moment map of each cloud. Although the rotation is not necessarily intrinsically solid body (i.e.\ the angular velocity may vary with radius within each cloud), the parameter \omeobs\ defined below nevertheless provides a useful single quantity to quantify the bulk rotation of individual clouds:
\begin{equation}
  \bar{v}(x,y)=ax+by+c\,\,\,,
\end{equation}
where $a$ and $b$ are the projected velocity gradient along the $x$- and the $y$-axis on the sky, respectively, and $c$ is a zero point,that we determine using the Interactive Data Language code \textsc{mpfit2dfun} \citep[][]{mpfit2dfun_ref}. We can thus calculate the projected angular velocity \omeobs\ and position angle of the rotation axis \rotaxis:
\begin{equation}
  \omega_{\mathrm{obs}}=\sqrt{a^2+b^2}
\end{equation}
and
\begin{equation}
  \phi_{\mathrm{rot}}=\tan^{-1}(b/a)\,\,\,,
\end{equation}
where the uncertainties of \omeobs\ and \rotaxis\ are estimated from the uncertainties of the parameters $a$ and $b$ using standard error propagation rules.

\autoref{tab:table1} lists the best-fitting \omeobs\ and \rotaxis. The projected velocity gradients \omeobs\ of the $170$ resolved clouds of NGC~5806 range from $0.01$ to $0.67$~\kms~pc$^{-1}$, with an average and median gradient of $0.10$ and $0.08$~\kms~pc$^{-1}$, respectively. These gradients are similar to those of the clouds in the MW ($\approx0.1$~\kms~pc$^{-1}$; \citealt{blitz1993,phillips1999,imarablitz2011}), M~33 ($\leq0.15$~\kms~pc$^{-1}$; \citealt{rosolowsky2003,imara2011,braine_2018_m33}), M~31 ($0$ -- $0.2$~\kms~pc$^{-1}$; \citealt{rosolowsky2007}) and M~51 ($\leq0.2$~\kms~pc$^{-1}$; \citealt{braine_2020_m51}), but they are smaller than those of the clouds in the ETGs NGC~4526 ($0.02$ -- $1.1$~\kms~pc$^{-1}$; \citealt{utomo2015}) and NGC~4429 ($0.05$ -- $0.91$~\kms~pc$^{-1}$; \citealt{liu2021ngc4429}).

\subsection{Origin of velocity gradients}
\label{sec:origins}

\begin{figure}
  \includegraphics[width=\columnwidth]{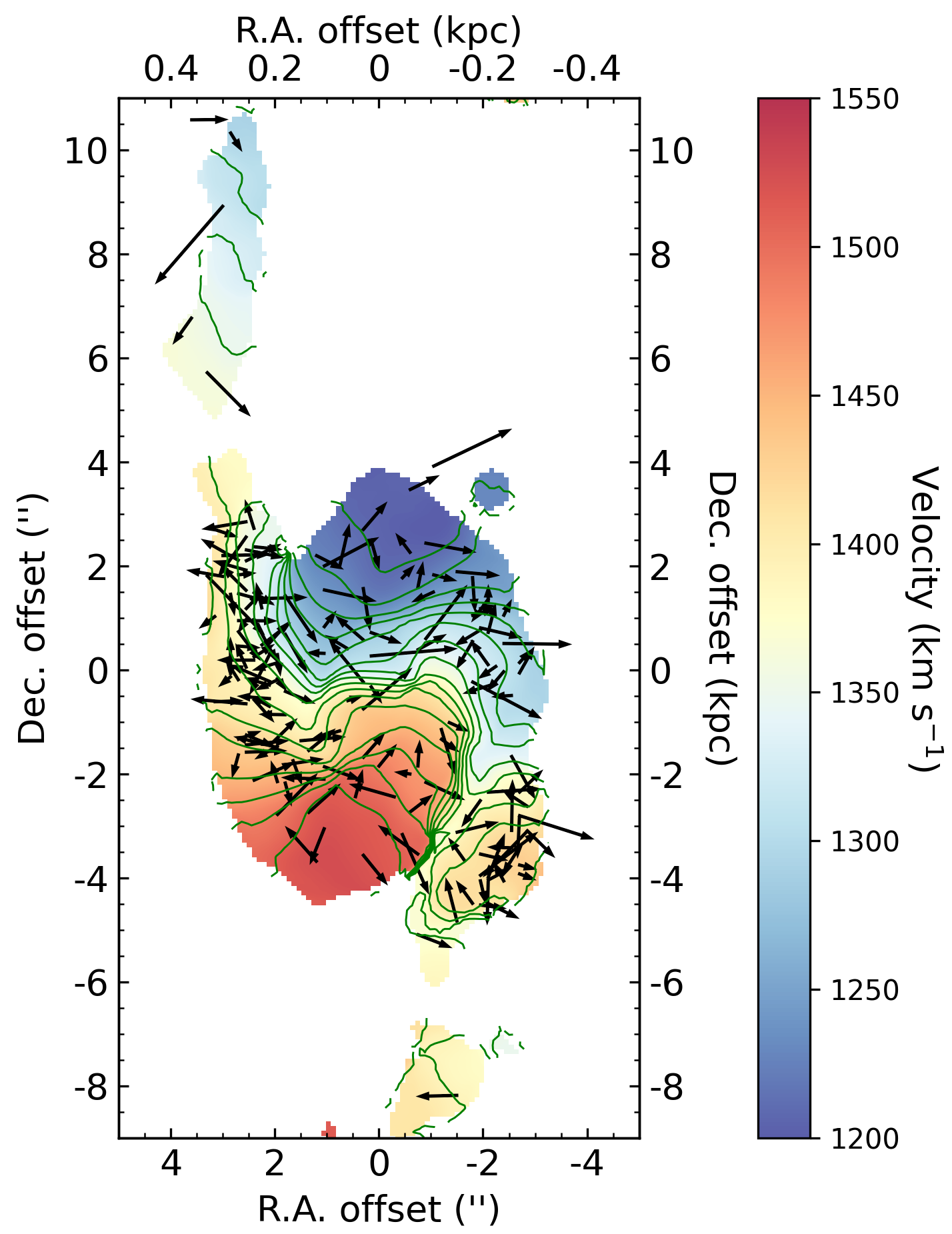}
  \caption{Projected angular momentum vectors of individual resolved GMCs in NGC~5806 (black arrows), overlaid on the \CO\ velocity map and isovelocity contours (green contours). The arrow length is proportional to the angular velocity \omeobs\ of each clouds.}
  \label{fig:velocity_map}
\end{figure}

To investigate the origin of the velocity gradients of the GMCs of NGC~5806, we first compare the velocity map of NGC~5806 to the projected rotation axes of its clouds. In \autoref{fig:velocity_map}, the projected rotation axes of the clouds (black arrows) are overlaid on the \CO\ mean velocity map and the isovelocity contours (green contours). The arrow length is proportional to the projected angular velocity of each cloud. If the rotation axes of the clouds are aligned with the galaxy isovelocity contours, the bulk rotation of the clouds is likely governed by the large-scale galaxy rotation. Conversely, if the rotation axes of clouds are randomly distributed, the bulk rotation of the clouds likely originates from other mechanisms, such as turbulence and/or cloud-cloud collisions (e.g. \citealt{burkert2000,wu2018_collision_turbulence,Li2018_ccc}), that perturb angular momentum conservation.

As shown in \autoref{fig:velocity_map}, the rotation axes of the of NGC~5806 clouds are not well aligned with the isovelocity contours, suggesting that the galaxy rotation does not affect the internal rotation of the clouds. This is similar to the case of the MW \citep{koda2006}, M~31 \citep{rosolowsky2007} and NGC~5064 \citep{liu_2023}, but different from that of the ETGs NGC~4526 \citep{utomo2015} and NGC~4429 \citep{liu2021ngc4429}, where the rotation axes are well aligned with the isovelocity contours.

\begin{figure*}
  \includegraphics[width=1.6\columnwidth]{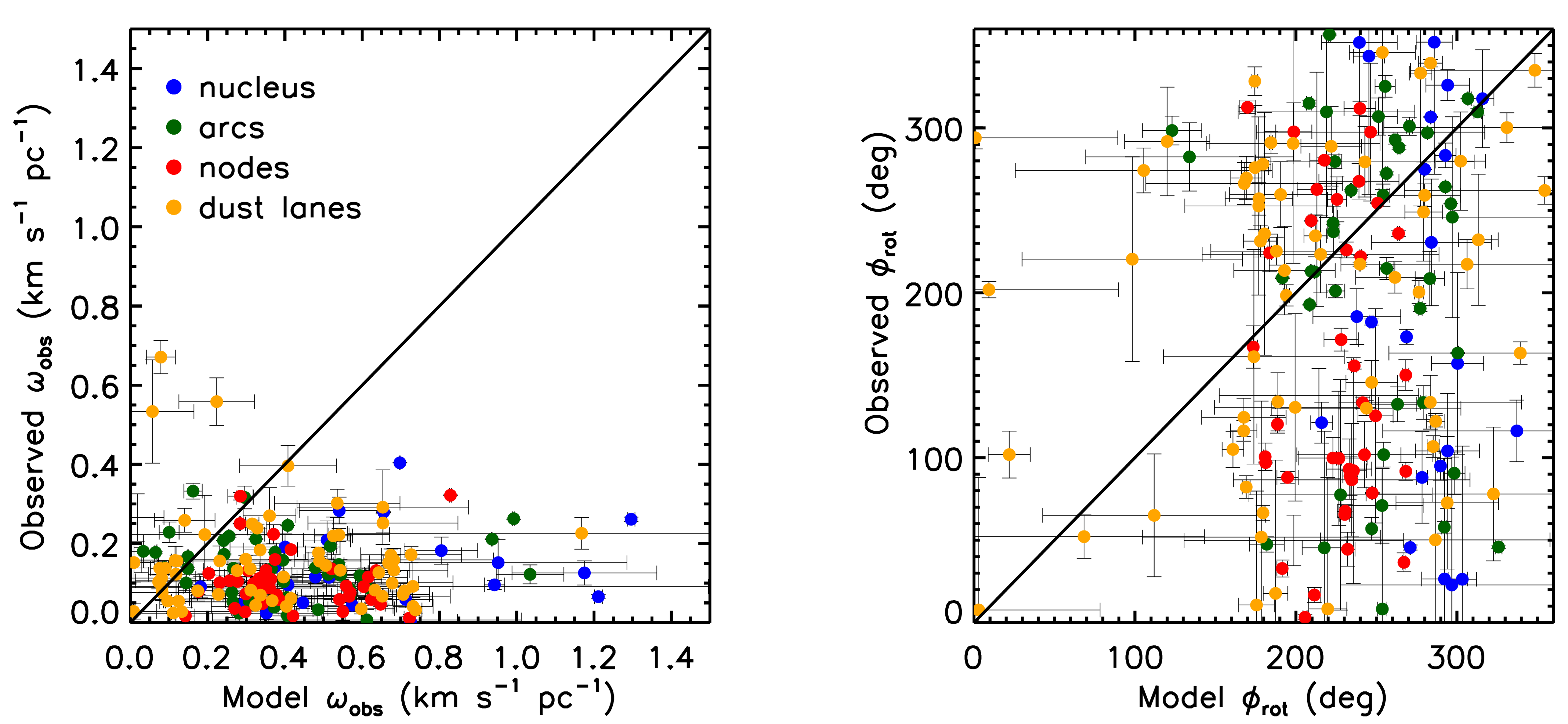}
  \centering
  \caption{Correlations between the modelled and observed projected angular velocities \omeobs\ (left) and position angles of the rotation axes \rotaxis\ (right) for the $170$ resolved clouds of NGC~5806. The data points are colour-coded by region and the black diagonal lines indicate the $1:1$ relations.}
  \label{fig:omega_model_comparison}
\end{figure*}

To further investigate this, we compare in \autoref{fig:omega_model_comparison} the measured angular velocities (\omeobs) and rotation axes (\rotaxis) of the clouds with those expected ($\omega_{\mathrm{model}}$ and $\phi_{\mathrm{model}}$), as calculated from a low-resolution (i.e.\ coarse-grained) \CO\ velocity map over the same position and area as each resolved cloud and using the same method as in \autoref{sec:gradients}.

The modelled angular velocities ($\omega_{\mathrm{model}}$) are on average $\approx3.5$ times larger than the observed ones. Furthermore, there is no clear correlation between the modelled and observed orientation of the cloud rotation axes. Although not easily visible in \autoref{fig:omega_model_comparison}, there are different trends across the different regions of NGC~5806. About half of the clouds ($13/25$) in the arcs have a small difference between the modelled and observed rotation axis orientations ($|\phi_{\mathrm{rot}}-\phi_{\mathrm{model}}|\leq50^{\circ}$), while only about one third of the clouds ($50/145$) in the other regions have such a small difference. Consequently, the velocity gradient of the clouds in the arcs are more likely to be governed by the large-scale galaxy rotation, presumably because the molecular gas there is less affected by the surrounding environment (e.g.\ AGN feedback and shocks) than the gas in other regions. Conversely, the velocity gradient of the clouds in the nucleus, nodes and dust lanes are more likely to be due to other origins (e.g.\ random turbulent motions and/or cloud-cloud collisions).

\citet{burkert2000} showed that the apparent rotation of clouds can arise from the clouds' turbulence. They claimed a relation of the form $\left(\frac{\omega}{\mathrm{km~s^{-1}~pc^{-1}}}\right)=1.6\left(\frac{R_{\mathrm{c}}}{\mathrm{0.1~pc}}\right)^{-1/2}$. This formulation yields $\omega=0.092$ ($0.086$) \kms~pc$^{-1}$ for the median (mean) cloud radius of $30.6$ ($34.5$) pc in NGC~5806. These are comparable to the median (mean) of our measured angular velocities, $0.10$ ($0.12$) \kms~pc$^{-1}$. It is thus most likely that the observed velocity gradients of the clouds of NGC~5806 are due to turbulent motions. However, we find that not all clouds have the same trend. The median cloud radii of the clouds in the nucleus, arcs, nodes and dust lanes are $27.4$, $37.8$, $35.8$ and $26.9$~pc, respectively, yielding expected angular velocities of $0.097$, $0.082$, $0.085$ and $0.098$~\kms~pc$^{-1}$, compared to median measured angular velocities of $0.095$, $0.13$, $0.07$ and $0.11$~\kms~pc$^{-1}$. The clouds in the arcs thus show a much larger deviation ($\approx45\%$) than those in the other regions. This implies that additional mechanisms supporting and/or generating cloud rotation are required in the arcs.

\begin{figure}
  \includegraphics[width=\columnwidth]{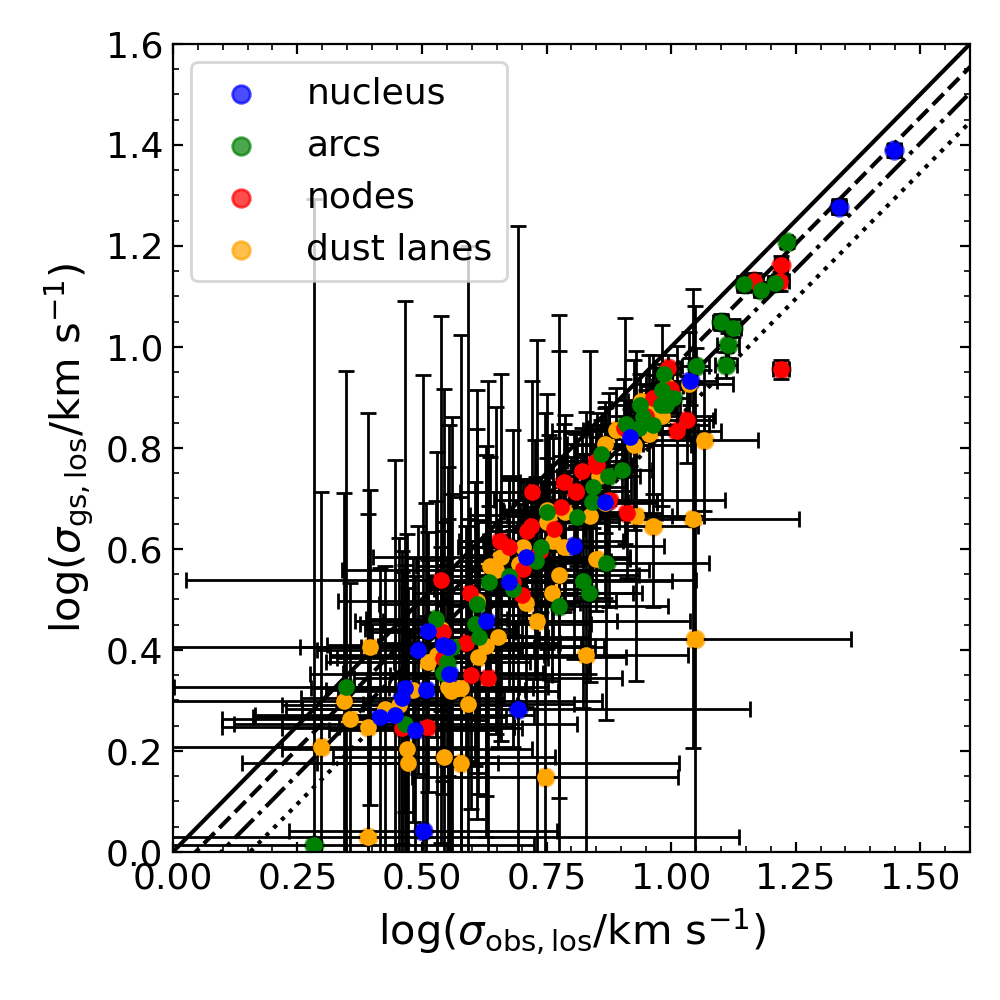}
  \caption{Comparison of our observed (\sigobs) and gradient-subtracted (\siggs) velocity dispersion measures for the $170$ resolved clouds of NGC~5806. The four black diagonal lines represent the $1:1$, $1:0.9$, $1:0.8$ and $1:0.7$ ratio, respectively.}
  \label{fig:sigma_compare}
\end{figure}

Another way to assess whether bulk motions due to galaxy rotation contribute significantly to the observed velocity dispersions and velocity gradients of the NGC~5806 clouds is to compare the observed velocity dispersions (\sigobs) with the gradient-subtracted velocity dispersions (\siggs), as shown in \autoref{fig:sigma_compare}. If the gradient-subtracted velocity dispersions are much smaller than the observed velocity dispersions, bulk motions are dominant in the clouds. More than half of the clouds ($107/170$) of NGC~5806 have a small difference between the two velocity dispersions (i.e.\ a ratio between the two velocity dispersions $\sigma_{\mathrm{obs,gs}}/\sigma_{\mathrm{obs,los}}>0.7$). Some clouds have somewhat larger deviations, but only four clouds have a difference of more than $5$~\kms. This further suggests that bulk motions due to galaxy rotation are not important to the NGC~5806 clouds. The observed velocity dispersions and velocity gradients are thus likely dominated by turbulence. In turn, we will use the observed velocity dispersions (\sigobs) rather than the gradient-subtracted velocity dispersions (\siggs) for the rest of our analyses, also consistent with previous GMC studies.

\section{Dynamical states of clouds}
\label{sec:dynamics}

Scaling relations (e.g.\ relations between the sizes, linewidths and luminosities of GMCs) have been used as a standard tool to investigate the dynamical states of clouds \citep[e.g.][]{larson1981,blitz2007}. Among them, the relation between the size and the linewidth (a.k.a.\ Larson's first relation) is generally considered the most fundamental. The size -- linewidth relation is known to have the form of a power law and is generally interpreted as the consequence of turbulent motions within clouds \citep[][]{falgarone1991,elmegreen1996,Lequeux2005}.

\begin{figure*}
  \centering
  \includegraphics[width=2\columnwidth]{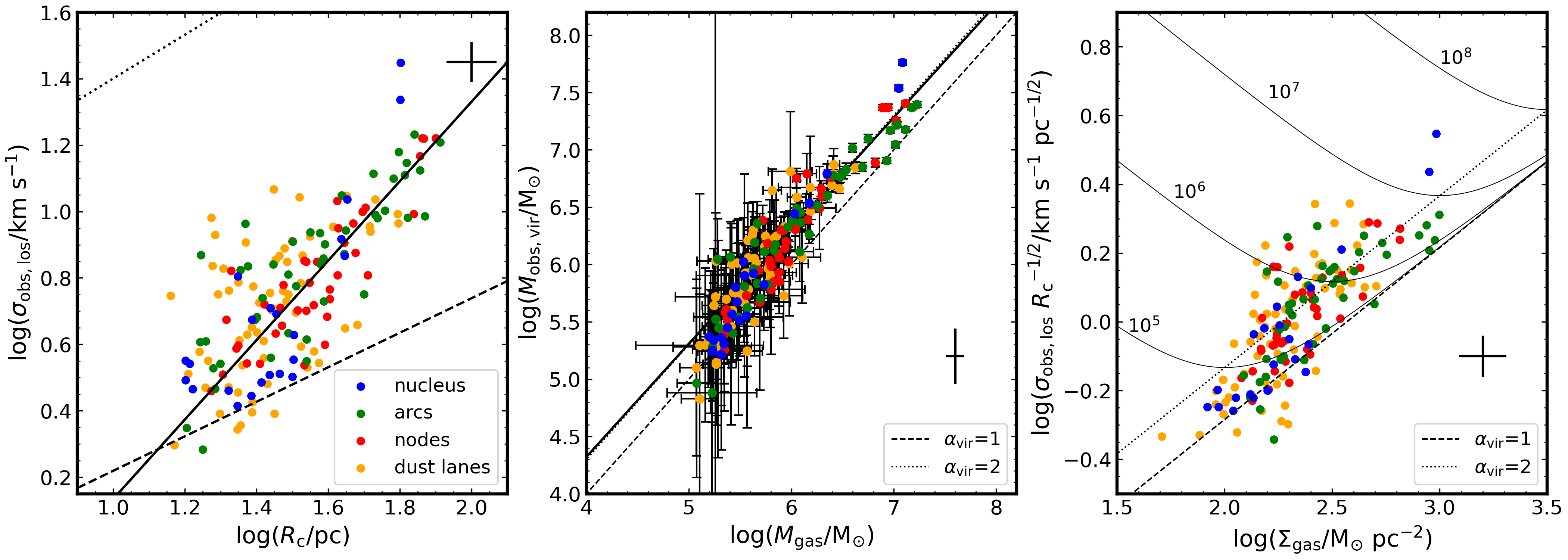}
  \caption{Left: size -- linewidth relation of the $170$ resolved clouds of NGC~5806. The black solid line shows the best-fitting power-law relation, while the black dashed and dotted lines show the relations for the clouds of the MW disc \citep{solomon1987} and CMZ \citep{kauffmann2017}, respectively. Middle: molecular gas mass -- virial mass relation for the same clouds. The black solid line shows the best-fitting power-law relation, while the black dashed and dotted lines indicate the $1:1$ and $2:1$ relations, respectively. Right: gas mass surface density -- $\sigma_{\mathrm{obs,los}}R_{\mathrm{c}}^{-1/2}$ relation for the same clouds. The black dashed and dotted diagonal lines show the solutions for a simple (i.e.\ $\alpha_{\mathrm{vir}}=1$) and a marginal (i.e.\ $\alpha_{\mathrm{vir}}=2$) virial equilibrium, respectively. The V-shaped black solid curves show solutions for pressure-bound clouds at different pressures ($P_{\mathrm{ext}}/k_{\mathrm{B}}=10^5$, $10^6$, $10^7$ and $10^8$~K~cm$^{-3}$). Data points are colour-coded by region in all three panels. Typical uncertainties are shown as a black cross in the top-right or bottom-right corner of each panel.}
  \label{fig:larson_obs}
\end{figure*}

The left panel of \autoref{fig:larson_obs} shows the size -- linewidth relation of all resolved clouds of NGC~5806, with the best-fitting power-law relation overlaid (black solid line), as well as that of the MW disc (black dashed line; \citealt{solomon1987}) and CMZ (black dotted line; e.g.\ \citealt{kauffmann2017}). There is a strong correlation between size and linewidth, with a Spearman rank correlation coefficient of $0.70$ and \textit{p}-value of $10^{-35}$.

Our best-fitting power law has a steep slope,
\begin{equation}
  \log\left(\frac{\sigma_{\mathrm{obs,los}}}{\mathrm{km\ s^{-1}}}\right)=(1.20\pm0.10)\log\left(\frac{R_{\mathrm{c}}}{\mathrm{pc}}\right)-1.07\pm0.16\,\,\,,
\end{equation}
with no clear difference between different regions. To achieve the best-fitting line with both \Rc\ and \sigobs\ errors, we use a hierarchical Bayesian model called \textsc{linmix} \citep[][]{kelly2007_linmix}.\footnote{Python version of the linmix algorithm has been provided by J. Meyers (\url{https://github.com/jmeyers314/linmix}).} This slope is steeper than that of the clouds in the MW disc ($0.5\pm0.05$; \citealt{solomon1987}) and the CMZ ($0.66\pm0.18$; \citealt{kauffmann2017}), but the zero-point ($0.09$~\kms) is much smaller than that of the CMZ ($5.5\pm1.0$~\kms; \citealt{kauffmann2017}). This slope is also much steeper than that of the clouds in M~31 ($0.7\pm0.2$; \citealt{rosolowsky2007}), M~33 ($0.45\pm0.02$; \citealt{rosolowsky2003}), NGC~4429 ($0.82\pm0.13$; \citealt{liu2021ngc4429}) and local dwarf galaxies ($0.60\pm0.10$; \citealt{bolatto2008}). Although the GMCs of barred-spiral galaxies have been investigated (e.g.\ M83 and NGC~1300; \citealt{hirota2018_m83_bar,maeda2020a_n1300_bar}), only the LMC shows a clear size -- linewidth relation, with a slope of $0.8$ \citep{wong2011LMC}.

Another scaling relation used to assess the dynamical states of clouds is the correlation between virial (\Mvir) and gas (\Mgas) mass. In the absence of non-gravitational forces, this quantifies the dynamical state of clouds based on the virial theorem. The virial parameter
\begin{equation}
  \label{eq:virial_parameter}
  \begin{split}
    \alpha_{\mathrm{vir}} & \equiv\frac{M_{\mathrm{vir}}}{M_{\mathrm{gas}}}=\frac{\sigma_{\mathrm{obs,los}}^2R_{\mathrm{c}}/b_{\mathrm{s}}G}{M_{\mathrm{gas}}}\\
    & =\frac{3M_{\mathrm{gas}}\sigma_{\mathrm{obs,los}}^2}{3b_{\mathrm{s}}GM_{\mathrm{gas}}^2/R_{\mathrm{c}}}=\frac{2K}{|U|}\,\,\,,
  \end{split}
\end{equation}
where $b_{\mathrm{s}}$ is a geometrical factor that quantifies the effects of inhomogeneities and/or non-sphericity of a cloud on its self-gravitational energy ($U$) and $K$ is the kinetic energy of random motions of the cloud. Here we adopt $b_{\mathrm{s}}=1/5$ assuming the clouds have homogeneous spherical shapes. If $\alpha_{\mathrm{vir}}\approx1$, a cloud is considered to be in virial equilibrium and is gravitationally bound, while if $\alpha_{\mathrm{vir}}\approx2$, the cloud is only marginally gravitationally bound. If $\alpha_{\mathrm{vir}}<1$, the cloud is likely to collapse gravitationally, while if $\alpha_{\mathrm{vir}}\gg1$, the cloud is either confined by non-gravitational forces (e.g.\ external pressure and/or magnetic fields) or it is short-lived (i.e.\ transient).

The middle panel of \autoref{fig:larson_obs} shows the virial masses of the resolved clouds of NGC~5806 (calculated using the observed velocity dispersion \sigobs) as a function of their gas masses, overlaid with the best-fitting power law (black solid line). The black dashed and dotted lines indicate the $\alpha_{\mathrm{vir}}=1$ and $\alpha_{\mathrm{vir}}=2$ relations, respectively. The best-fitting power law estimated from the linmix algorithm is
\begin{equation}
  \log\left(\frac{M_{\mathrm{obs,vir}}}{\mathrm{M_{\odot}}}\right)=(0.99\pm0.03)\log\left(\frac{M_{\mathrm{gas}}}{\mathrm{M_{\odot}}}\right)+0.38\pm0.20\,\,\,,
\end{equation} 
implying that the resolved clouds of NGC~5806 are virialised on average. Similarly to Larson's first relation in the left panel, the resolved clouds in the different regions tend to have the same best-fitting slope, but the clouds in the arcs are slightly more massive than those in the other regions.

\begin{figure}
  \includegraphics[width=\columnwidth]{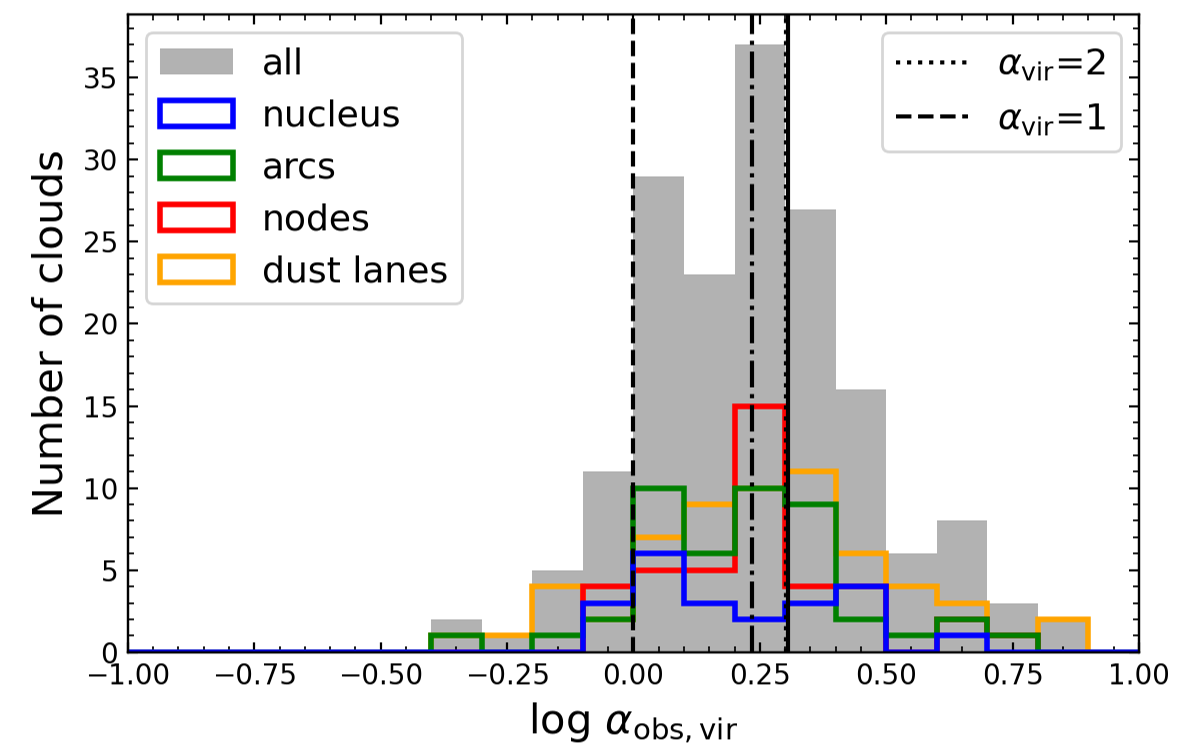}
  \caption{Distribution of $\log(\alpha_{\mathrm{obs,vir}}$) of all the resolved clouds of NGC~5806 (grey histogram) and only the clouds in each of the four different regions (coloured histograms). The black solid and dot-dashed lines show the mean and median of the distribution, respectively, while the black dashed and dotted lines indicate $\alpha_{\mathrm{vir}}=1$ and $\alpha_{\mathrm{vir}}=2$, respectively.}
  \label{fig:alpha_obs}
\end{figure}

To investigate the virialisation of the resolved clouds of NGC~5806 further, we also explore the distribution of \virialpha\ for the whole galaxy and each region individually, as shown in \autoref{fig:alpha_obs}. The mean (median) of $\alpha_{\mathrm{vir}}$ is $2.02$ ($1.72$), indicating that on average the clouds are marginally bound. However, \virialpha\ has a broad distribution and only about half of the clouds ($89/170$) lie between $\alpha_{\mathrm{vir}}=1$ and $\alpha_{\mathrm{vir}}=2$. About $40\%$ of the clouds ($62/170$) have $\alpha_{\mathrm{vir}}>2$, while only a few clouds ($19/170$) have $\alpha_{\mathrm{vir}}<1$. Unlike other physical quantities such as size, gas mass, velocity dispersion and gas mass surface density (see \autoref{sec:properties}), there is no significant difference across the different regions.

Lastly, we consider the correlation between gas mass surface density ($\Sigma_{\mathrm{gas}}$) and $\sigma_{\mathrm{obs,los}}R_{\mathrm{c}}^{-1/2}$, providing yet another constraint on the physics of clouds \citep{field2011}. Regardless of how well clouds obey Larson's first relation, if the clouds are virialised, the clouds should be clustered around $\sigma_{\mathrm{obs,los}}R_{\mathrm{c}}^{-1/2}=\sqrt{{\pi\alpha_{\mathrm{vir}}Gb_{\mathrm{s}}\Sigma_{\mathrm{gas}}}}$, as shown by the black dashed ($\alpha_{\mathrm{vir}}=1$) and dotted ($\alpha_{\mathrm{vir}}=2$) diagonal lines in the right panel of \autoref{fig:larson_obs}. If clouds are not virialised ($\alpha_{\mathrm{vir}}\gg1$), external pressure ($P_{\mathrm{ext}}$) should play an important role to constrain the clouds (or the clouds are likely transient structures). In this case, clouds will be distributed around the black solid V-shape curves in the right panel of \autoref{fig:larson_obs}:
\begin{equation}
  \sigma_{\mathrm{obs,los}}R_{\mathrm{c}}^{-1/2}=\sqrt{\frac{\pi\alpha_{\mathrm{vir}}G\Sigma_{\mathrm{gas}}}{5}+\frac{4\,P_{\mathrm{ext}}}{3\,\Sigma_{\mathrm{gas}}}}
\end{equation}\label{eq:pressure_virial}
\citep{field2011}.

The right panel of \autoref{fig:larson_obs} shows the relation between $\sigma_{\mathrm{obs,los}}R_{\mathrm{c}}^{-1/2}$ and \surfgas\ for all the resolved clouds of NGC~5806, showing that they are broadly distributed. The gas mass surface densities vary by $1.5$ orders of magnitude and reveal a positive correlation with $\sigma_{\mathrm{obs,los}}R_{\mathrm{c}}^{-1/2}$. Given the uncertainties, some clouds with $\alpha_{\mathrm{vir}}>2$ distributed across the V-shaped curves do seem to be bound by high external pressures ($P_{\mathrm{ext}}/k_{\mathrm{B}}\gtrsim10^5$~\punit, if indeed they are bound). In particular, two clouds in the nucleus at very high pressures ($P_{\mathrm{ext}}/k_{\mathrm{B}}\gtrsim10^7$~\punit) might be affected by nuclear activity. In addition, as expected from the right panel of \autoref{fig:larson_obs} (but not explicitly shown in the figure), there is a strong correlation between \sigobs\ and \surfgas\ (Spearman rank correlation coefficient of $0.73$ and \textit{p}-value of $2\times10^{-35}$), while the correlation between \Rc\ and \surfgas\ is much weaker (Spearman rank correlation coefficient of $0.38$ and \textit{p}-value of $7\times10^{-12}$).

In summary, \autoref{fig:larson_obs} shows that the size -- linewidth relation of the resolved clouds of NGC~5806 has a slope that is twice as steep as that of MW disc clouds, while most of the clouds are only marginally bound ($\langle\alpha_{\mathrm{vir}}\rangle\approx2$).

\section{Discussion}
\label{sec:discussion}

\subsection{Turbulence maintained by bar-driven gas inflows}
\label{sec:discussion_turbulence}

High velocity dispersions are present in the central regions of NGC~5806, especially in the nodes (up to 60~\kms; see top-right panel of \autoref{fig:moments}). Individual clouds also have large velocity widths, and the clouds have a very steep size-linewidth relation and relatively large virial parameters (see \autoref{sec:dynamics}). Could all these facts be due to the large-scale bar of NGC~5806?

Recent observations of barred spiral galaxies have shown that bars can drive gas inflows and contribute to the high velocity dispersions observed in the central regions of many galaxies. For example, \citet{salak2016} reported high molecular gas velocity dispersions ($\gtrsim40$~\kms) in the node regions of NGC~1808 ((R)SAB(s)a), that are due to gas streaming along the bar toward the nuclear ring. \citet{sato2021} also reported high gas velocity dispersions ($\gtrsim40$~\kms) in the nuclear ring of NGC~613, especially at the interface between the nuclear ring and the large-scale bar (i.e.\ the node regions) where gas inflows are observed.

\begin{figure*}
  \centering
  \includegraphics[width=2\columnwidth]{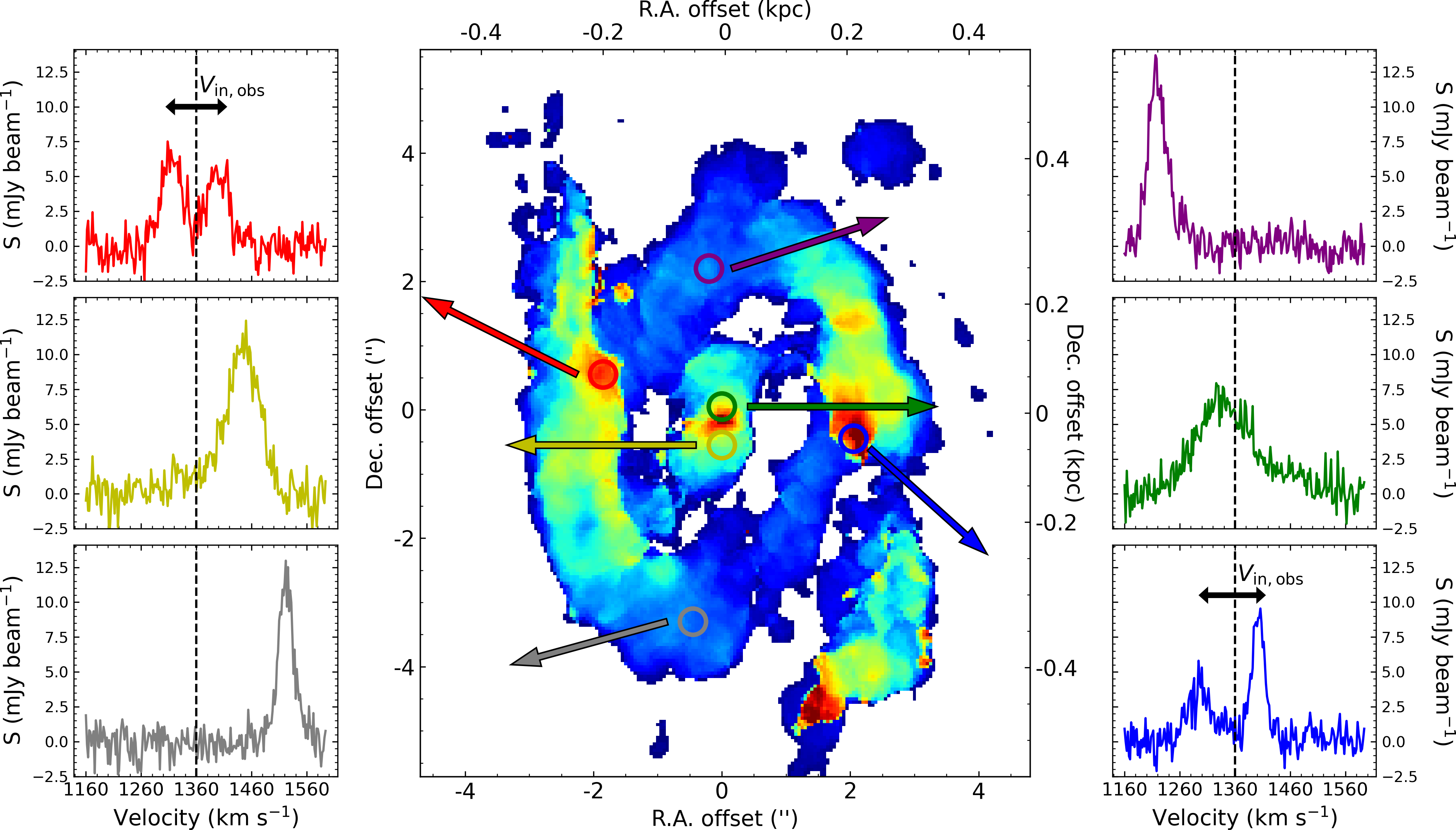}
  \caption{Velocity dispersion map of NGC~5806. The red and blue circles indicate regions where velocity dispersions are above $50$~\kms\ in the nodes, the yellow and green circles regions where velocity dispersions are above $50$~\kms\ in the nucleus, and the purple and grey circles regions where the velocity dispersions are below $20$~\kms\ in the arcs. Inset panels show the corresponding spectra and the black vertical dashed line in each panel indicates the systemic velocity of NGC~5806. For double peaked spectra, the black arrows indicate the velocity differences $V_{\mathrm{in,obs}}$ discussed in the text.}
  \label{fig:spectrum}
\end{figure*}

To investigate whether the bar in NGC~5806 also drives gas inflows, we probe the shapes of the \CO\ line-of-sight (LOS) velocity distributions (LOSVDs) across the different regions, and illustrate specific trends in \autoref{fig:spectrum}. High velocity dispersions are present in the nuclear ring, especially in the nodes (up to $60$~\kms). The LOSVDs in the nodes also have shapes that are totally different from those in the other regions: the LOSVDs in the nodes are often double peaked within a single synthesised beam (see the red and blue circles in \autoref{fig:spectrum}), while the LOSVDs in the rest of the nuclear ring (i.e.\ the arcs) have only narrower single peaks (see the purple and grey circles in \autoref{fig:spectrum}). The LOSVDs in the nucleus have broad and skewed shapes with single peaks (see the yellow and green circles in \autoref{fig:spectrum}), likely due to strong shear and/or AGN feedback (and beam smearing), that can render the molecular gas more turbulent \citep[e.g.][]{wada2009}.

The double peaks of the node LOSVDs imply that there are multiple clouds (at least two) along each LOS. Furthermore, systematically, for each double peak in the nodes, one peak smoothly connects to the LOSVD of the nearest dust lane, the other to the LOSVD of the nuclear ring, suggesting that the molecular gas in the dust lanes flows toward the nuclear ring. Molecular gas thus appears to be streaming along the bi-symmetric offset dust lanes, causing collisions in the molecular gas in the nodes. This is analogous to the situation in the MW, where \citet{kruijssen2014} suggested that gas inflows along the bar may be responsible for driving the turbulence in the CMZ. More generally, collisions and shocks resulting from gas streaming into a nuclear ring (from the large-scale bar) can cause significant random motions in the gas \citep[e.g.][]{kruijssen2014,federrath2016,sormani2019,salas2021,wallace2022}.

To measure the relative velocity between the gas inflowing along the offset dust lanes and the nuclear ring, we consider individual LOSVDs in the nodes and estimate the velocity difference between the two dominant peaks. The measured average velocity difference is $V_{\mathrm{in,obs}}\approx100$~\kms\ in both the eastern and the western node. The gas inflow velocity is then $V_{\mathrm{in}}=V_{\mathrm{in,obs}}/\sin i\approx120$~\kms\ \citep[e.g.][]{sato2021}. Adopting this relative velocity, the total mass inflow rate along the two dust lanes can be estimated as
\begin{equation}
  \label{eq:mass_inflow_rate}
  \begin{split}
    \dot{M}_{\mathrm{in}} & =2\langle\Sigma_{\mathrm{gas}}\rangle W_{\mathrm{in}}V_{\mathrm{in}}\\
    & \approx5~{\mathrm{M_\odot~yr^{-1}}}\,\,\,,\\
  \end{split}
\end{equation}
where the width of the gas inflow $W_{\mathrm{in}}$ to each node is taken to be $\approx100$~pc and the mean molecular gas mass surface density in the dust lanes is $\langle\Sigma_{\mathrm{gas}}\rangle\approx200$~\Msun~pc$^{-2}$. Similarly, to estimate the contribution of the gas inflows in driving the turbulence, we can estimate the total kinetic energy per unit time provided by the gas inflows as
\begin{equation}
  \label{eq:energy_inflow_rate}
  \begin{split}
    \dot{E}_{\mathrm{in}} & \approx\frac{1}{2}\dot{M}_{\mathrm{in}}V_{\mathrm{in}}^2\\
    & \approx3.5\times10^4~{\mathrm{M_\odot~km^2~s^{-2}~yr^{-1}}}\,\,\,.\\
    \end{split}
\end{equation}

If turbulence in the nuclear ring is indeed maintained by the bar-driven gas inflows, the turbulence energy dissipation per unit time $\dot{E}_{\mathrm{diss}}$ should be balanced by the input kinetic energy per unit time, i.e.\ $\dot{E}_{\mathrm{diss}}\approx\dot{E}_{\mathrm{in}}$. The energy per unit time dissipated by the observed turbulence can be estimated as
\begin{equation}
  \label{eq:energy_dissipation_rate}
  \begin{split}
    \dot{E}_{\mathrm{diss}} & \approx M_{\mathrm{NR}}\langle\sigma_{\mathrm{NR}}\rangle^3/(2\,h_{\mathrm{NR}})\\
    & \approx2.8\times10^4~{\mathrm{M_\odot~km^2~s^{-2}~yr^{-1}}}\,\,\,.\\
    \end{split}
\end{equation}
\citep[e.g.][]{maclow_klessen2004}, where $M_{\mathrm{NR}}\approx2.2\times10^8$~M$_\odot$, $\langle\sigma_{\mathrm{NR}}\rangle\approx16$~km~s$^{-1}$ and $h_{\mathrm{NR}}\approx16$~pc are the total mass, mean velocity dispersion and scale height of the molecular gas in the nuclear ring, respectively. This is indeed approximately equal to our estimated $\dot{E}_{\mathrm{in}}$, so bar-driven molecular gas inflows are a viable mechanism to explain the high velocity dispersions present in the nuclear ring.

The aforementioned scale height was estimated as $h_{\mathrm{NR}}=\langle\sigma_{\mathrm{NR}}\rangle/\kappa_{\mathrm{NR}}$ \citep{lin_pringle1987}, where $\kappa_{\mathrm{NR}}$ is the epicyclic frequency at the nuclear ring radius, that can be calculated as $\kappa^2_{\mathrm{NR}}\equiv\left.\left(R\frac{d\Omega^2(R)}{dR}+4\Omega^2(R)\right)\right|_{R=R_{\mathrm{NR}}}$, where $\Omega(R)\equiv V_{\mathrm{c}}(R)/R$, $V_{\mathrm{c}}(R)$ is the circular velocity of NGC~5806 and $R_{\mathrm{NR}}$ is the radius (at the centre) of the nuclear ring ($R_{\mathrm{NR}}\approx330$~pc). As the molecular gas in NGC~5806 is so dynamically cold, we took $V_{\mathrm{c}}(R)$ to be the observed rotation curve, derived from our data cube using \textsc{3dbarolo} \citep{Teodoro2015_3dbarolo} and from our first-moment map using \textsc{2dbat} \citep{oh2018_2dbat}. Both approaches yield consistent results, leading to $\kappa_{\mathrm{NR}}\approx1$~km~s$^{-1}$~pc$^{-1}$ and thus the adopted scale height $h_{\mathrm{NR}}\approx16$~pc.


\subsection{Nuclear ring GMC lifetimes}
\label{sec:discussion_gmc_ring}

\begin{figure*}
  \includegraphics[width=2\columnwidth]{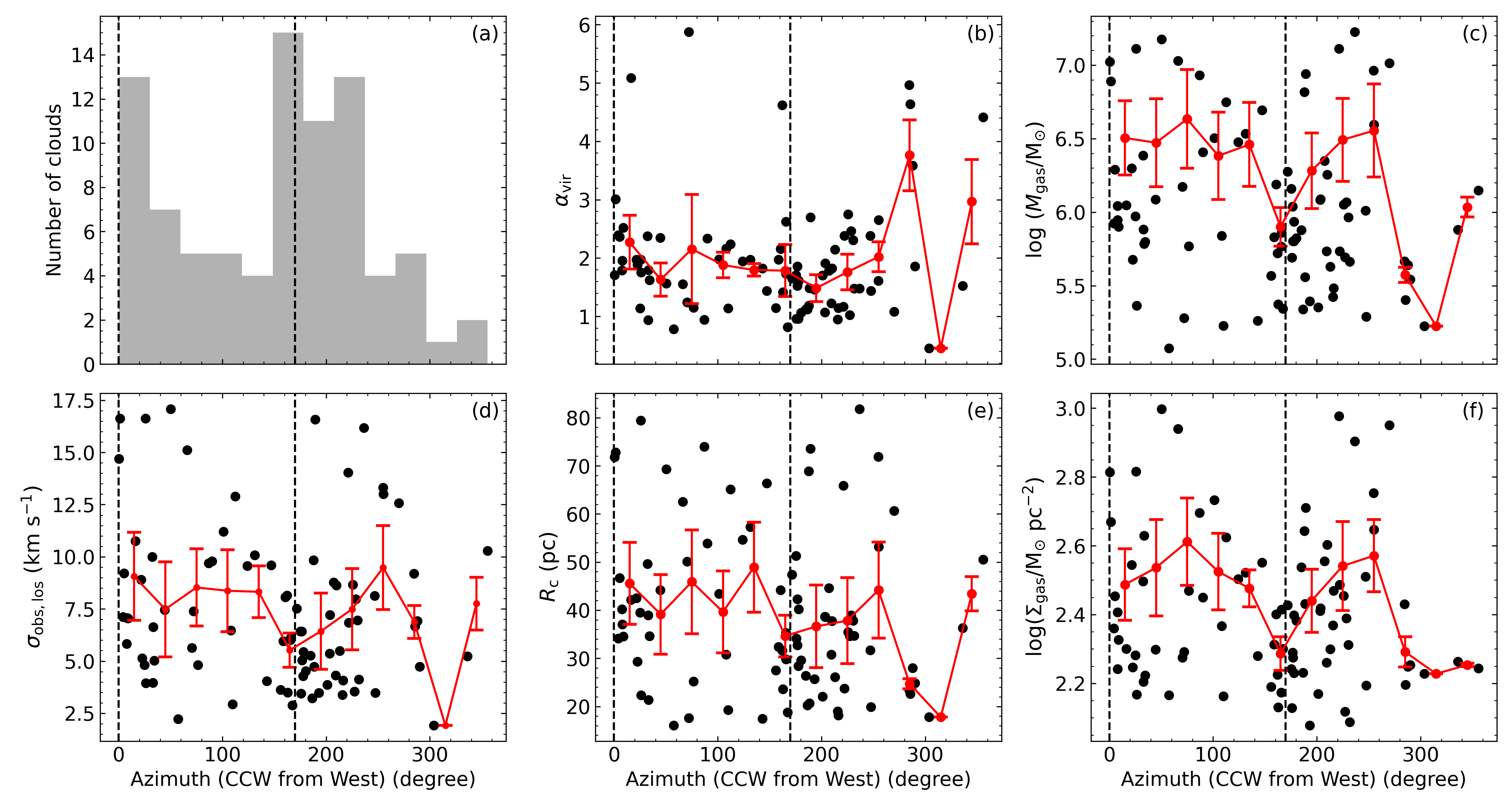}
  \caption{GMC properties in the nuclear ring (both arcs and nodes), as a function of the azimuthal angle (measured counter-clockwise from the western node). From left to right, top to bottom: number of resolved clouds, virial parameter, gas mass, velocity dispersion, size and gas mass surface density. The red data points are averages in radial bins of width $30\degr$, while the red error bars indicate the $1\sigma$ scatter within each radial bin. Black vertical dashed lines indicate the positions of the two nodes.}
  \label{fig:ring_cloud}
\end{figure*}

As argued above (\autoref{sec:discussion_turbulence}), bar-driven gas inflows should strongly influence the cloud properties in the nuclear ring (see also \citealt{salak2016,sato2021}). It is thus important to probe whether cloud properties vary azimuthally along the ring. \autoref{fig:ring_cloud} shows both the number of clouds and a number of cloud properties (virial parameter, gas mass, velocity dispersion, size and gas mass surface density) as a function of azimuthal angle (measured counter-clockwise from the western node). Interestingly, while none of the cloud properties varies significantly with azimuthal angle (see panels~(b) -- (f) of \autoref{fig:ring_cloud}), the number of clouds (as well as the CO surface brightness) strongly decreases from one node to the other (see panel~(a) of \autoref{fig:ring_cloud}).

\begin{figure} 
  \centering
  \includegraphics[width=0.7\columnwidth]{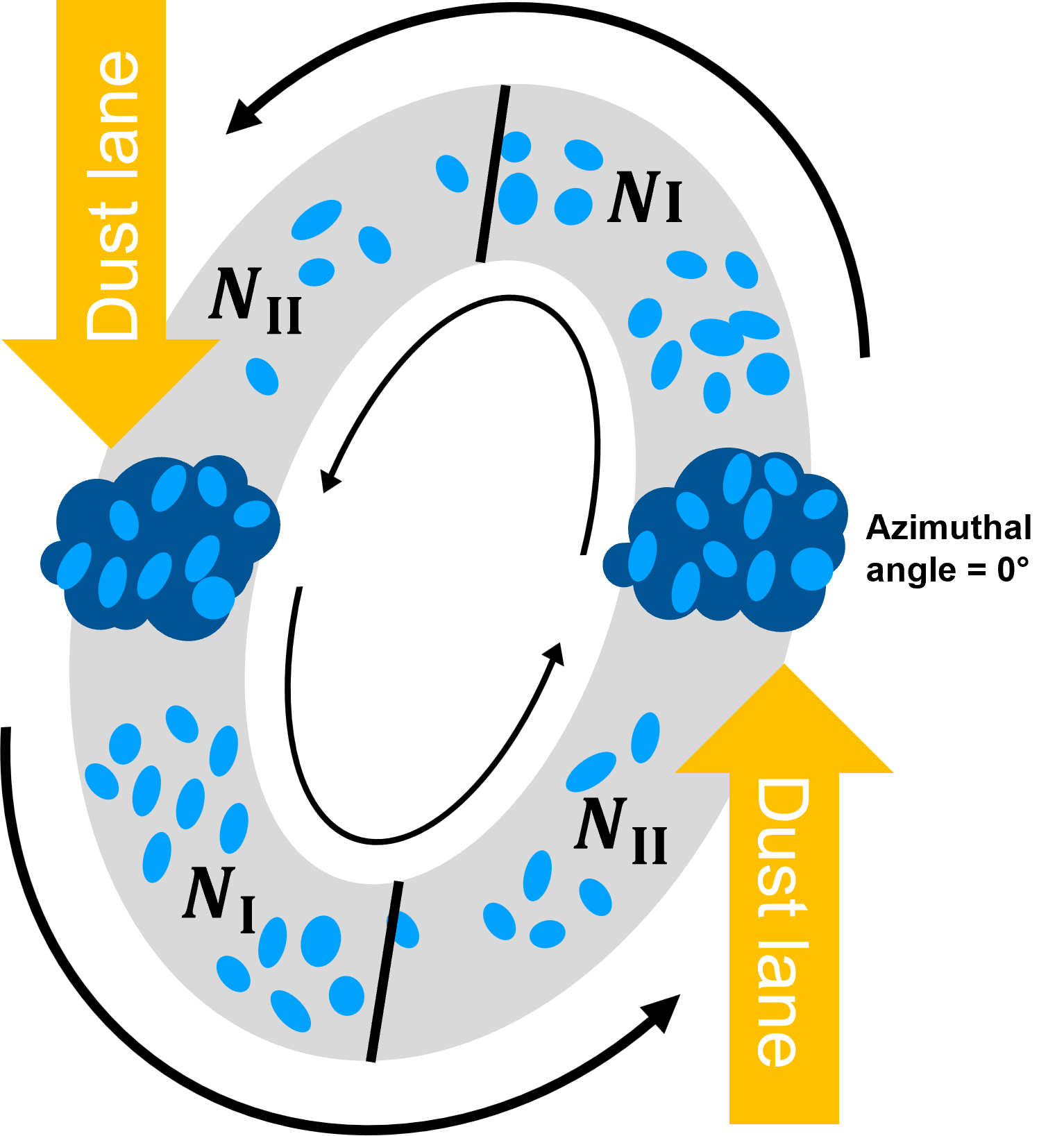}
  \caption{Schematic diagram of the scenario envisaged for the nuclear ring of NGC~5806. The nulear ring is shown as a large pale grey annulus, clouds as small blue filled ellipses within the nuclar ring, and the two offset dust lanes as thick yellow arrows. The two solid vertical black lines indicate the midpoints that divide each half of the nuclear ring into two zones. $N_{\mathrm{I}}$ and $N_{\mathrm{II}}$ are the number of clouds in each of those two zones.}
  \label{fig:ring_cartoon}
\end{figure}

Which mechanisms can cause this steep decrease of the cloud number along the nuclear ring downstream from the nodes? Most likely, when molecular gas from the large-scale bar enters the nuclear ring at the nodes, the ensuing violent collisions will lead to the formation of many clouds. For their number to decrease, these clouds formed at the nodes must then be gradually destroyed while moving along the nuclear ring (see \autoref{fig:ring_cartoon}). This may be due to a number of mechanisms such as further cloud-cloud collisions, shear, stellar feedback, AGN feedback and/or violent turbulence.

Irrespective of the exact cloud disruption mechanism, the observed azimuthal variation of the cloud number embodies the resulting cloud lifetimes. Indeed, the characteristic cloud lifetime can be estimated from the travel time $t_{\mathrm{travel}}$ between the two nodes and the fraction of clouds lost as they move $F_{\mathrm{lost}}$ (i.e.\ the decline of the number of clouds as their travel between the two nodes):
\begin{equation}
  \label{eq:lifetime}
  t_{\mathrm{lifetime}}=\frac{t_{\mathrm{travel}}}{2}\frac{1}{F_{\mathrm{lost}}}\,\,\,.
\end{equation}
We note that this method to estimate the cloud lifetimes in the nuclear ring of a barred galaxy is similar to that introduced by \citet{meidt2015} to estimate the cloud lifetimes in the inter-arm region of a spiral galaxy. However, as noted below, we measure the travel time $t_{\mathrm{travel}}$ with respect to the large-scale bar rotating pattern, as inspired by the work of \citet{koda2021}.

We can estimate $t_{\mathrm{travel}}$ as
\begin{equation}
  \label{eq:travel_time}
  t_{\mathrm{travel}}=\pi R_{\mathrm{NR}}/(V_{\mathrm{c,NR}}-\Omega_{\mathrm{p}}R_{\mathrm{NR}})
\end{equation}
\citep[see also][]{koda2021}, where $V_{\mathrm{c,NR}}$ is the circular velocity at the (radius of the) centre of nuclear ring ($V_{\mathrm{c,NR}}\approx150$~\kms) and $\Omega_{\mathrm{p}}$ is the pattern speed of the large-scale bar. The pattern speed of the bar of NGC~5806 has never been measured. We could obtain a firm lower limit to the travel time by adopting $\Omega_{\mathrm{p}}=0$ in \autoref{eq:travel_time} (yielding $t_{\mathrm{travel}}\gtrsim6.9$~Myr), but instead we estimate the bar pattern speed by assuming its corotation radius is located at $1.2\pm0.2$ times the half-length of the bar, as is the case for most barred disc galaxies \citep[e.g.][]{anthanassoula1992_dust_shock,aguerri1998_bar}. \citet{erwin_2005_bar} measured the deprojected half-length of the bar of NGC~5806 to be $38\arcsec$ or $3.9$~kpc at our adopted distance, leading to a pattern speed $\Omega_{\mathrm{p}}=V_{\mathrm{c}}[(1.2\pm0.2)\,R_{\mathrm{bar}}]/[(1.2\pm0.2)\,R_{\mathrm{bar}}]=45_{-7}^{+11}$~km~s$^{-1}$~kpc$^{-1}$. In turn, using \autoref{eq:travel_time}, this leads to a travel time $t_{\mathrm{travel}}=7.4_{-0.1}^{+0.3}$~Myr.

To estimate $F_{\mathrm{lost}}$, we first measure the number of clouds $N_{\mathrm{I}}$ and $N_{\mathrm{II}}$ in two adjacent zones (each mirrored on both halves of the nuclear ring) that span equal ranges of azimuth, as shown in \autoref{fig:ring_cartoon} (see also Figure~1 of \citealt{meidt2015}). The fraction of clouds lost between the two nodes is
\begin{equation}
  \label{eq:Flost}
  F_{\mathrm{lost}}=\frac{N_{\mathrm{I}}-N_{\mathrm{II}}}{N_{\mathrm{I}}}\,\,\,.
\end{equation}
Counting the numbers of clouds in the first half of the nuclear ring (from the western node to the eastern node) yields $N_{\mathrm{I}}=23$ and $N_{\mathrm{II}}=9$ (and thus $F_{\mathrm{lost}}=0.61$), while in the second half (from the eastern node to the western node) this yields $N_{\mathrm{I}}=40$ and $N_{\mathrm{II}}=11$ (and thus $F_{\mathrm{lost}}=0.72$). This apparent loss of clouds along the nuclear ring is probably tightly related to the decrease of the CO intensity downstream from the nodes (see the top-left panel of \autoref{fig:moments}).

Combined with our estimated travel time, these two fractions of lost clouds yield two cloud lifetime estimates, that we take as a range $t_{\mathrm{lifetime}}=5.1$ -- $6.3$~Myr. This cloud lifetime is smaller than that of clouds in the central $3.5$~kpc radius of M~51 ($20$ -- $50$~Myr; \citealt{meidt2015}), nearby galaxies ($10$ -- $100$~Myr; \citealt{jeffreson2018,chevance2020}), the LMC ($\approx11$~Myr; \citealt{ward2022_lmc_lifetime}) and between spiral arms in disc galaxies ($\approx100$~Myr; e.g.\ \citealt{koda2009_gmc_lifetime}), but it is larger than that of clouds in the CMZ ($1$ -- $4$~Myr; e.g.\ \citealt[][]{kruijssen2015_cmz_lifetime,jeffreson2018_cmz}).

\subsection{Nuclear ring GMC destruction mechanisms}
\label{sec:discussion_gmc_destruction}

Having estimated the lifetimes of the clouds in the nuclear ring of NGC~5806, we now briefly discuss the possible mechanisms regulating those lifetimes. A cloud's lifetime is mainly set by cloud-cloud collisions, shear, stellar feedback, AGN feedback and/or violent turbulence \citep[e.g.][]{meidt2015,jeffreson2018,chevance2020,kim2022_gmc}. We therefore now derive the relevant time scales of these processes, and compare them with our derived cloud lifetime. Processes that take longer than the estimated cloud lifetime are likely to play a less important role setting the cloud lifetime than processes with shorter timescales.

{\bf Cloud-cloud collisions.} Cloud-cloud collisions can be an important mechanism limiting cloud lifetimes, as clouds can be destroyed when merging with other clouds. The cloud-cloud collision timescale in the nuclear ring can be estimated as $t_{\mathrm{coll}}=1/N_{\mathrm{mc}}D_{\mathrm{c}}\sigma_{\mathrm{cc}}$ \citep{koda2006}, where $N_{\mathrm{mc}}$ is the cloud number surface density in the nuclear ring, $D_{\mathrm{c}}$ is the mean cloud diameter in the nuclear ring ($2\,\langle R_{\mathrm{c}}\rangle\approx78$~pc; see panel~(e) of \autoref{fig:ring_cloud}) and $\sigma_{\mathrm{cc}}$ is the cloud-cloud velocity dispersion, generally assumed to be $\approx10$~\kms\ \citep[e.g.][]{koda2006,inutsuka2015_cloud}. To estimate $N_{\mathrm{mc}}$, we consider the $85$ nuclear ring clouds contained within an elliptical annulus of inner semi-major axis length $230$~pc, outer semi-major axis length $370$~pc and ellipticity $0.3$, that nicely encloses the nuclear ring, yielding $N_{\mathrm{mc}}\approx450$~kpc$^{-2}$ and in turn $t_{\mathrm{coll}}\approx3.1$~Myr. Our derived collision timescale is approximately half the estimated cloud travel time between the nodes $t_{\mathrm{travel}}$ and is smaller than the estimated cloud lifetime $t_{\mathrm{lifetime}}$.

{\bf Shear.} Shear generally appears to be an important mechanism regulating cloud lifetimes in galaxy centres, where strong shear can lead to mass loss and/or complete cloud dispersal \citep[e.g.][]{meidt2015,jeffreson2018}. We estimate the shear timescale as $t_{\mathrm{shear}}=1/2A$ \citep{liu2021ngc4429}, where $A\equiv\frac{1}{2}\left(\frac{V_{\mathrm{c}}(R_{\mathrm{NR}})}{R_{\mathrm{NR}}}-\left.\frac{dV_{\mathrm{c}}}{dR}\right|_{R_{\mathrm{NR}}}\right)\approx0.15$~\kms~pc$^{-1}$ is Oort's constant evaluated at (centre of) the nuclear ring using the aforementioned rotation curve, yielding $t_{\mathrm{shear}}\approx3.2$~Myr. This is again approximately half the cloud travel time between the nodes and is smaller than the estimated cloud lifetime.

{\bf Stellar feedback.} The destruction of molecular clouds by stellar feedback occurs on a feedback timescale $t_{\mathrm{feedback}}$, i.e.\ the timescale of coexistence of molecular gas and stars within a cloud \citep[e.g.][]{chevance2020}. This can be estimated by measuring the spatial offset between cold gas (the fuel for star formation, traced by e.g.\ CO) and star formation (traced by e.g.\ H$\alpha$) through the now widely-used ``tuning fork'' diagram \citep{kruijssen_longmore2014,kruijssen2018}. However, the absence of a map of a star-formation tracer at both high angular resolution and free of dust extinction prohibits a direct measurement of the feedback timescale in NGC~5806. \citeauthor{chevance2020}'s (\citeyear{chevance2020}) molecular gas measurements in nine nearby star-forming disc galaxies range from $1$ to $5$~Myr, with a typical timescale $t_{\mathrm{feedback}}\approx3.5$~Myr that we adopt for the clouds in our galaxy.

{\bf AGN feedback.} Nuclear activity is a powerful mechanism that can severely affect the medium surrounding a nucleus. Several observational studies have reported that AGN feedback is the most likely mechanism to explain the high velocity dispersions of molecular gas (and even molecular gas disruption) in galaxy centres \citep[e.g.][]{schawinski2009_agn_molecule,simionescu2018_agn_molecule,nesvadba2021_agn_gmc}. Several simulations also support AGN having a significant impact on the molecular gas in galaxy centres \citep[e.g.][]{wada2009,mukherjee2018_jet_moleucle}. However, these studies have also shown that this impact on the surrounding media is limited to several hundred parsecs in radius in the galactic discs (while extending beyond $1$~kpc perpendicularly to the discs). Furthermore, in NGC~5806, not only is the mm continuum not detected in our observations (\autoref{sec:data}), but only the nucleus (inner $100$~pc radius) was classified as AGN/shocks by \citet{errozferrer2019} using \citeauthor{bpt1981}'s (\citeyear{bpt1981}) diagnostics and optical integral-field spectroscopic observations. All these results thus suggest that the AGN of NGC~5806 is unlikely to directly affect the molecular gas in the nuclear ring.

{\bf Turbulence.} Strong turbulence could be another important process dispersing clouds in the nuclear ring \citep[e.g.][]{dobbs_pettitt2015,kim2022_gmc}. Its effect can be characterised by a cloud's turbulent crossing timescale, $t_{\mathrm{cross}}\approx 2R_{\mathrm{c}}/\sigma_{\mathrm{obs,los}}$ \citep[e.g.][]{kruijssen2019}. We can thus estimate the turbulent crossing timescales of the clouds in the nuclear ring, yielding timescales from $5$ to $20$~Myr, with a mean turbulent crossing timescale $\langle t_{\mathrm{cross}}\rangle\approx11$~Myr. This is much larger than our estimated cloud travel time between the nodes and our estimated cloud lifetime.


Overall, we can rule out turbulence as an important factor limiting the cloud lifetimes in the nuclear ring, as it acts on timescales much longer than the characteristic lifetime of the clouds ($t_{\mathrm{cross}}\approx 11$~Myr while $t_{\mathrm{lifetime}}\approx5.1$ -- $6.3$~Myr). This is consistent with the fact that the nuclear ring clouds have a mean virial parameter of $2.02$ and only $\approx30\%$ ($26/85$) of the clouds have $\alpha_{\mathrm{vir}}>2$ (see \autoref{sec:dynamics} and \autoref{fig:alpha_obs}). On the other hand, the timescales estimated for cloud-cloud collisions, shear and stellar feedback are comparable to each other, relatively short ($\approx3$ -- $3.5$~Myr) and all smaller than the estimated cloud lifetime, implying they could all play an important role setting cloud lifetimes. 

\subsection{Steep size -- linewidth relation}
\label{sec:discussion_steep_larson}

\begin{figure}
  \includegraphics[width=\columnwidth]{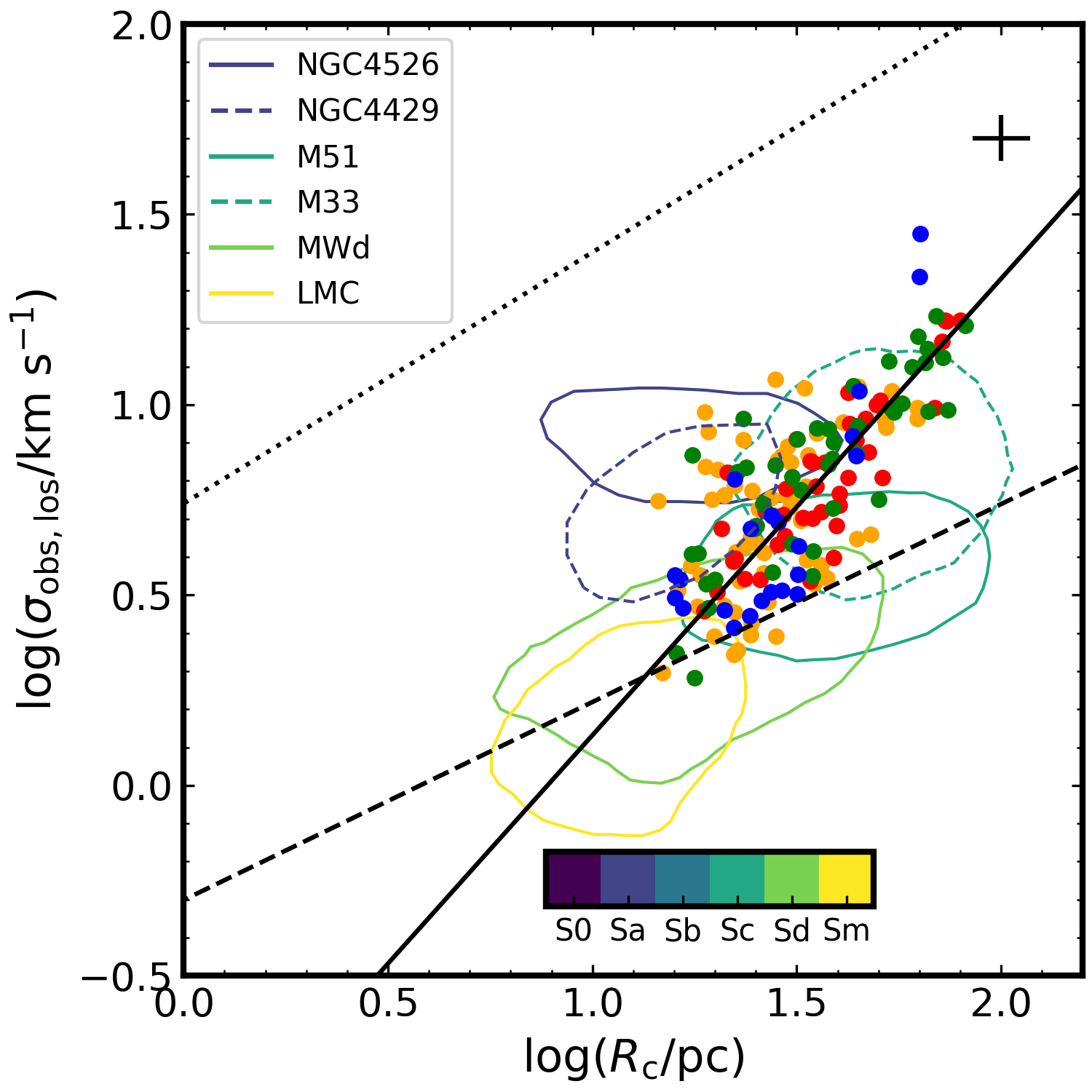}
  \caption{Size -- linewidth relation of extragalactic GMCs. Coloured circles show the resolved clouds of NGC~5806, while coloured contours encompass $68\%$ of the distribution of the data points for each galaxy (NGC~4526, \citealt{utomo2015}; NGC~4429, \citealt{liu2021ngc4429}; M~51, \citealt{colombo2014m51}; M~33, \citealt{gratier2012M33}; MWd, \citealt{heyer2009}; LMC, \citealt{wong2011LMC}). Contours are colour-coded by galaxy morphological type. The black solid line shows the best-fitting power-law relation of all NGC~5806 resolved clouds, the black dashed line that of the MWd clouds \citep{solomon1987} and the black dotted line that of the CMZ clouds \citep[][]{kauffmann2017}.}
  \label{fig:larson_only}
\end{figure}

As observations with different spatial resolutions are likely to trace different cloud sizes, we compare in \autoref{fig:larson_only} the size -- linewidth relations of NGC~5806 (spatial resolution $\approx24$~pc and sensitivity $\sigma_{\mathrm{rms}}\approx0.8$~K), the LMC ($\approx11$~pc and $\approx0.3$~K) and the two ETGs NGC~4526 ($\approx20$~pc and $\approx0.7$~K) and NGC~4429 ($\approx13$~pc and $\approx0.5$~K), whose observations have spatial resolutions and sensitivities similar to each other. While the observations of the MWd \citep{heyer2009}, M~51 \citep{colombo2014m51} and M~33 \citep{gratier2012M33} are more different, we also include them in \autoref{fig:larson_only} as those galaxies have a morphological type more similar to that of NGC~5806.

As discussed in \autoref{sec:dynamics}, the GMCs of NGC~5806 have a size -- linewidth relation with a power-law slope of $1.20\pm0.10$, much steeper than those found in other galaxies (e.g.\ $0.5$ -- $0.7$ in the MW, \citealt{solomon1987,kauffmann2017}; $0.4$ -- $0.8$ in nearby galaxies, \citealt{rosolowsky2003,rosolowsky2007,bolatto2008,wong2011LMC}).

The steep slope of the size -- linewidth relation of NGC~5806 is unlikely to be due primarily to the fact that it is measured in the central region of the galaxy (as opposed to the galaxy disc), as the size -- linewidth relations measured in the centres of other WISDOM galaxies appear to be much shallower and more similar to that of the MWd (e.g.\ no correlation in NGC~4526, \citealt{utomo2015}; slope of $0.6\pm0.1$ in NGC~5064, \citealt{liu_2023}; slope of $0.3\pm0.07$ in NGC~1387, \citealt{liang_2023}). A possible exception in the current WISDOM sample is the ETG NGC~4429, for which the GMCs in the central kpc-size disc have a size -- linewidth slope of $0.82\pm0.13$. However, once contamination of the cloud velocity dispersions by large-scale galaxy rotation (inducing bulk rotation within the clouds) is removed, the NGC~4429 clouds have a shallower slope than that of MWd clouds ($\approx0.24$; \citealt{liu2021ngc4429}).

A steep cloud size -- linewidth relation is also present in the central region of the WISDOM dwarf lenticular galaxy NGC~404 \citep{liu2022_ngc404}, although that study focused on much smaller structures, i.e\ clumps with sizes of $\approx3$~pc, so the comparison is arguably inappropriate. Overall, the distinct environment of galaxy centres is thus unlikely to be the only driver of the observed steep cloud size -- linewidth relation of NGC~5806.

The steep slope of the size -- linewidth relation is thus more likely to be due to gas inflows and shocks induced by the large-scale bar of NGC~5806. That bar-driven gas inflows contribute to the high velocity dispersions in the nuclear ring of NGC~5806 has already been discussed in \autoref{sec:discussion_turbulence}. However, the bar also drives strong shocks in the central region, as illustrated by the offset dust lanes and associated moleculas gas (a generic prediction of bar-driven shocks; e.g.\ \citealt{anthanassoula1992_dust_shock}).

In NGC~5806, turbulence can not dissipate all the energy through increasingly small spatial scales (i.e.\ through the usual turbulent ``cascade''), as kinetic energy is also being spent on shocks and/or gas compression \citep[e.g.][]{maclow1999,maclow_klessen2004,cen2021}. Since the energy transmission is no longer conservative ($\sigma_{\mathrm{obs,los}}\propto R_{\mathrm{c}}^{1/3}$ for a constant mass energy density transfer rate, \citealt{kolmogorov1941}; $\sigma_{\mathrm{obs,los}}\propto R_{\mathrm{c}}^{3/5}$ for a constant volumetric energy density transfer rate, \citealt{cen2021}), the size -- linewidth relation slope is expected to be steeper than $1/3$ -- $3/5$, as is indeed the case. An analogous example is probably that of the CMZ. Indeed, the MW also has a large-scale bar and the CMZ is most likely the equivalent of a nuclear ring in a barred galaxy, and the CMZ cloud size -- linewidth relation is rather steep (slope of $0.66\pm0.18$; \citealt{kauffmann2017}), if not as steep as that of NGC~5806.

The LMC and NGC~4526 also each have a large-scale bar, but comparisons with those galaxies are not justified as the LMC has a (poorly understood) off-centred bar strongly affected by a tidal interaction \citep[e.g.][]{deVaucouleurs1972_lmc_bar,vanderMarel2001_LMC_bar} while NGC~4526 has a relatively weak bar \citep{buta2007_bar}. Further studies of the impact of bars on the size -- linewidth relation would be highly valuable.

\subsection{Dependence of the virial parameter on cloud properties}
\label{sec:virial_relation}

\begin{figure*}
  \centering
  \includegraphics[width=2\columnwidth]{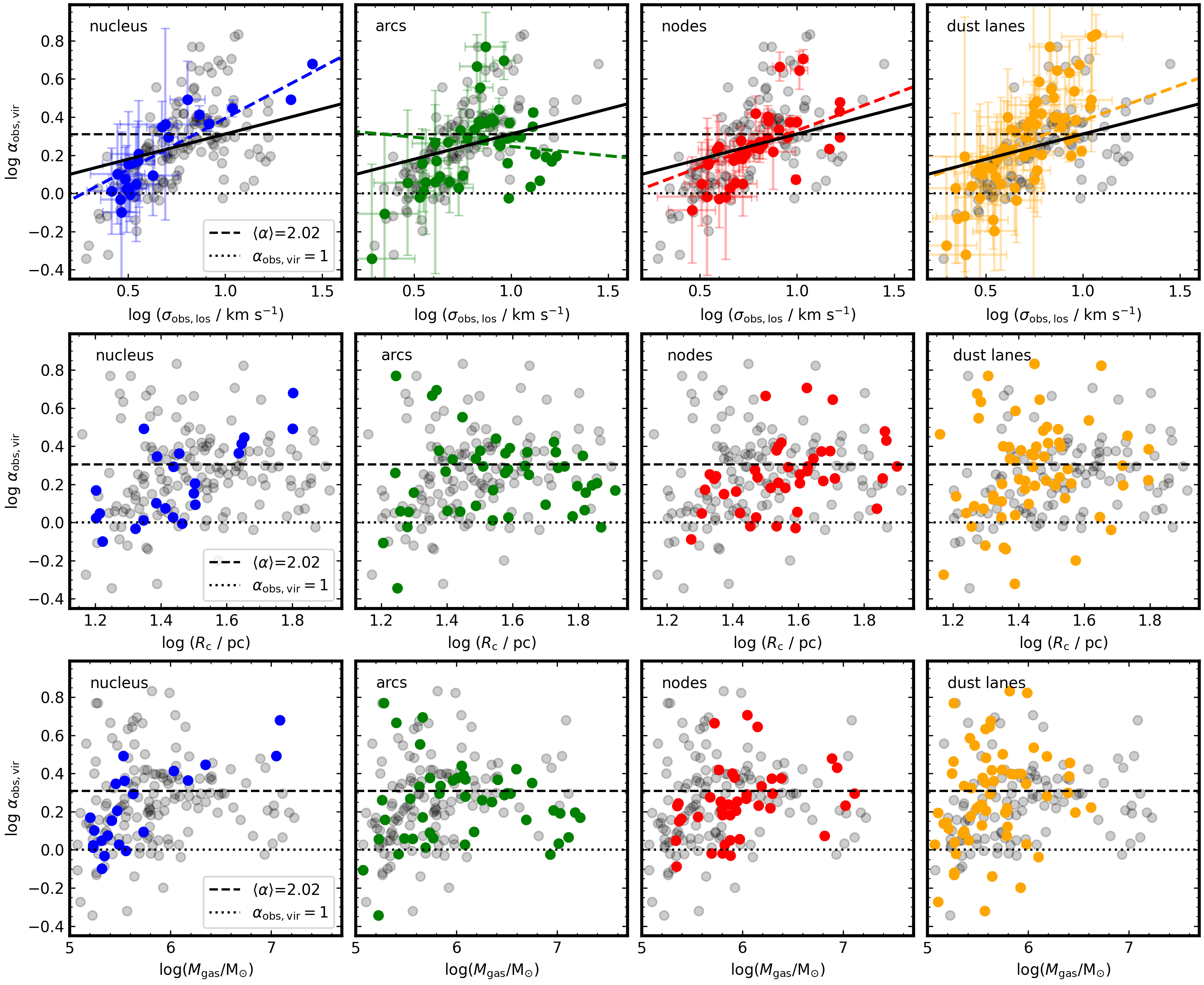}
  \caption{Dependence of the virial parameter (\virialpha) on the cloud velocity dispersion (\sigobs; first row), size (\Rc; second row) and gas mass (\Mgas; third row) for all the resolved clouds of NGC~5806. From left to right, the panels focus on the clouds in the nucleus (blue data points), arcs (green data points), nodes (red data points) and dust lanes (yellow data points); grey circles show all other resolved clouds. Black dotted lines indicate $\alpha_{\mathrm{vir}}=1$ and black dashed lines the mean virial parameter of all resolved clouds ($\langle\alpha_{\mathrm{vir}}\rangle=2.02$). In the first row, the black solid lines show the best-fitting power-law relation of all resolved clouds, while the coloured dashed line in each panel shows the best-fitting power-law relation of the resolved clouds in that region only.}
  \label{fig:virial_relation}
\end{figure*}

By definition ($\alpha_{\mathrm{vir}}\equiv\frac{\sigma_{\mathrm{obs,los}}^2R_{\mathrm{c}}}{b_{\mathrm{s}}GM_{\mathrm{gas}}}$; see \autoref{eq:virial_parameter}), assuming all quantities are independent, the virial parameter \virialpha\ is expected to have clear dependencies on the velocity dispersion (\sigobs), size (\Rc) and gas mass (\Mgas). However, for virialised clouds, these are expected to be correlated \citep[see e.g.][]{shetty2010}. To reveal which physical quantity primarily affects the virialisation of the clouds, we therefore probe the dependence of \virialpha\ on these quantities in \autoref{fig:virial_relation}.

There is no clear dependence of \virialpha\ on either \Rc\ or \Mgas\ for all the resolved clouds of NGC~5806 (second and third row of \autoref{fig:virial_relation}). This is inconsistent with previous studies that showed a clear negative correlation between \virialpha\ and \Mgas\ (e.g.\ \citealt{shetty2010,Miville2017b,veltchev2018}). The clouds of the nucleus do show correlations, but these are weak and largely depend on two exceptionally massive clouds. We also note that there is no correlation between \virialpha\ and \surfgas\ (Spearman rank correlation coefficient of $0.18$ and \textit{p}-value of $0.094$). On the other hand, there is a clear positive correlation between \virialpha\ and \sigobs\ in the nucleus, nodes and dust lanes, while the arcs shows a negative correlation (first row of \autoref{fig:virial_relation}), with a Spearman rank correlation coefficient of $0.67$ and \textit{p}-value of $5\times10^{-20}$ for all resolved clouds. The best-fitting power-law of all clouds estimated from the linmix algorithm is $\alpha_{\mathrm{vir}}\propto\sigma_{\mathrm{obs,los}}^{0.27\pm0.09}$ (black solid lines in the first row of \autoref{fig:virial_relation}). The trends of the node and dust lane clouds are similar to that of all the clouds, while the nucleus clouds have a steeper slope and the arc clouds have a negative correlation. Although the strength of the correlation between \virialpha\ and \sigobs\ varies between regions (Spearman coefficient (\textit{p}-value) of $0.89$ ($4\times10^{-8}$), $0.37$ (0.003), $0.77$ ($3\times10^{-6}$) and $0.78$ ($2\times10^{-13}$) for the nucleus, arcs, nodes and dust lanes, respectively), this result clearly shows that the gravitational boundedness of GMCs in NGC~5806 primarily depends on how turbulent they are (i.e.\ \sigobs) rather than other physical properties. As expected, the positive power-law index implies that clouds with weaker turbulence are more gravitationally bound than clouds with stronger turbulence. 

\subsection{CO conversion factor}
\label{sec:conversion_factor}

By rejecting the assumption of a uniform CO-to-H$_2$ conversion factor $X_{\mathrm{CO}}$ and assuming instead that all resolved clouds are virialised ($\alpha_{\mathrm{obs,los}}=1$), we can infer the variations of $X_{\mathrm{CO}}$.

\begin{figure}
  \includegraphics[width=\columnwidth]{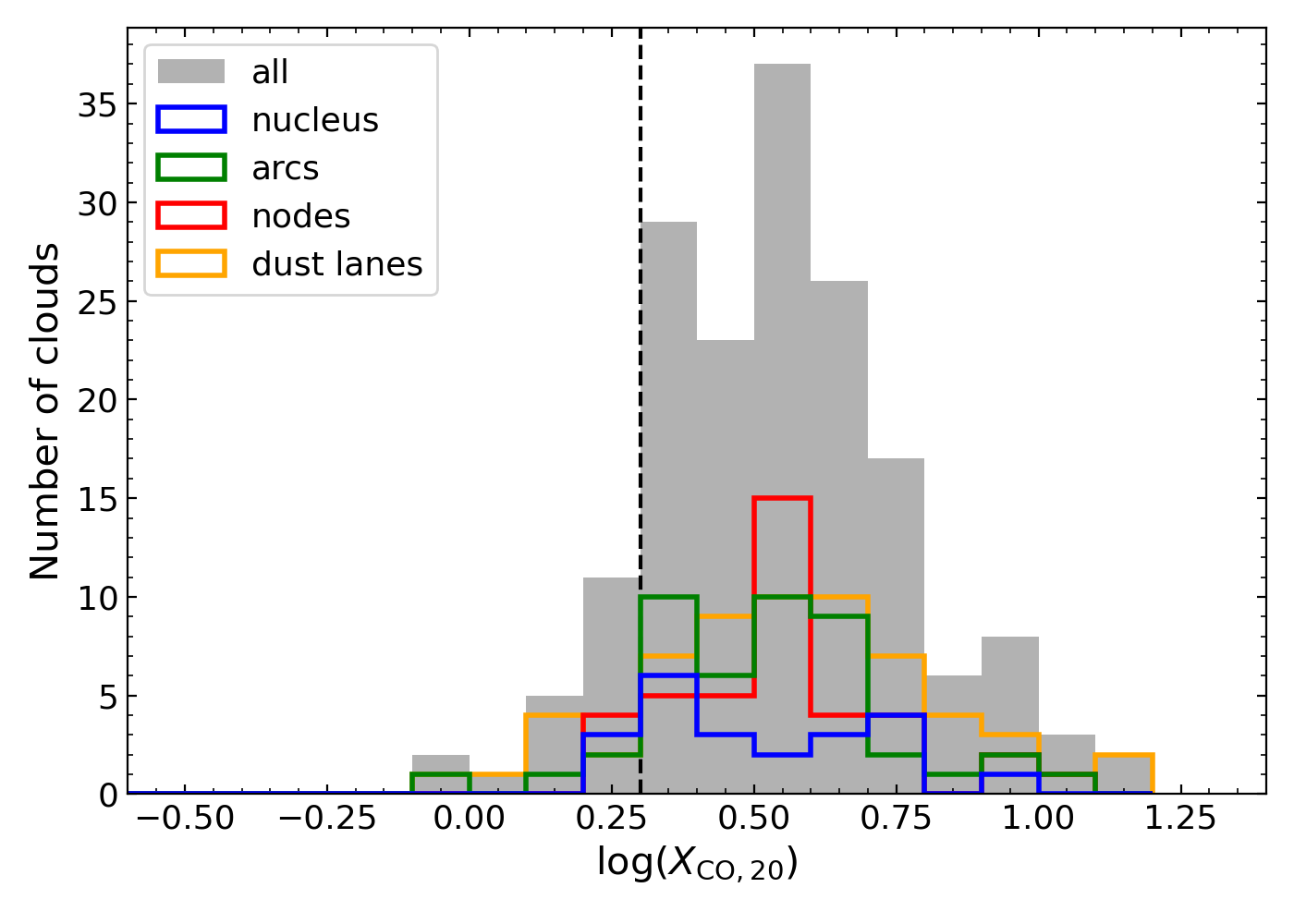}
  \caption{Distribution of $\log(X_{\mathrm{CO,20}})$ of all resolved clouds of NGC~5806 (grey histogram) and only clouds in each of the four different regions (coloured histograms), assuming all clouds are in virial equilibrium. The black dashed line indicates $X_{\mathrm{CO,20}}=2$ ($X_{\mathrm{CO}}=2\times10^{20}$~cm$^{-2}$~(K~km~s$^{-1}$)$^{-1}$).}
  \label{fig:Xco_obs}
\end{figure}

We define 
\begin{equation}
  X_{\mathrm{CO,20}}\equiv\frac{X_{\mathrm{CO}}}{1\times10^{20}~\mathrm{cm}^{-2}~(\mathrm{K~km~s}^{-1})^{-1}}\,\,\,,
\end{equation} 
and show in \autoref{fig:Xco_obs} the distribution of $X_{\mathrm{CO,20}}$ of all resolved clouds of NGC~5806 (identical to \autoref{fig:alpha_obs} within a scaling factor), with a logarithmic mean of $0.61$ ($X_{\mathrm{CO,20}}\approx4.05$). Considering that a typical $X_{\mathrm{CO,20}}$ for Milky Way disc clouds is $2$, the average conversion factor of NGC~5806 is slightly larger. It is also larger than that derived in the centres of many galaxies (i.e.\ $X_{\mathrm{CO,20}}\approx0.1$ -- $1$; \citealt{oka1998,israel2009,Sandstrom2013}). However, the mean conversion factor of NGC~5806 is comparable to those of $12$ nearby galaxies ($\approx3.5$; \citealt{bolatto2008}). If we applied the median $X_{\mathrm{CO,20}}$ from the literature ($\approx0.5$) to NGC~5806, the mean virial parameter of the clouds would be $4$ times higher, which seems unrealistically high. Finally, \autoref{fig:Xco_obs} shows that the distributions of $X_{\mathrm{CO,20}}$ in the four regions are similar to each other, with similar means. A spatially-varying conversion factor might therefore not be necessary in the central kiloparsec of NGC~5806.

It should of course be noted that we used the \CO\ transition (and assumed a constant $R_{21}=1$) instead of the $^{12}$CO(1-0) transition used by the majority of previous studies. Thus, a spatially-varying $R_{21}$ may be required to infer more plausible molecular gas masses. Observations of different $^{12}$CO transitions would help to derive more accurate $X_{\mathrm{CO}}$ and cloud properties.

\section{Conclusions}
\label{sec:conclusions}

We presented \CO\ ALMA observations of the barred spiral galaxy NGC~5806 at $25\times22$~pc$^2$ spatial resolution, and identified $366$ GMCs ($170$ of which are spatially and spectrally resolved) using our modified version of the \textsc{cpropstoo} code. The molecular gas of NGC~5806 has a highly structured distribution with a clear nucleus, nuclear ring (including nodes and arcs) and offset dust lanes. We studied the cloud properties and scaling relations in the different regions, and investigated how they are influenced by the large-scale bar.

The main findings are as follows:
\begin{enumerate}
\item The GMCs of NGC~5806 have slightly larger molecular gas masses ($10^5$ -- $10^{7.5}$~\Msun) and comparable sizes ($15$ -- $85$~pc) but larger velocity dispersions ($1.6$ -- $30$~\kms) and gas mass surface densities ($80$ -- $1000$~M$_\odot$~pc$^{-2}$) than those of MW disc and Local Group galaxy clouds. On the other hand, they have larger sizes and gas masses but smaller velocity dispersions and gas mass surface densities than those of CMZ clouds (\autoref{fig:histogram}). The GMCs in the nuclear ring are larger, brighter and more turbulent than the clouds in the nucleus, while the GMCs in the dust lanes are intermediate.
    
\item The cumulative gas mass function of the NGC~5806 clouds follows a truncated power law with a slope of $-1.72\pm0.12$. The nodes and arcs (i.e.\ the nuclear ring) have cloud mass functions that are significantly shallower than those of the nucleus and dust lanes, suggesting at least two different GMC populations, and massive GMCs are preferentially located in the nuclear ring. (\autoref{fig:mass_spectrum}).
    
\item The GMCs of NGC~5806 have a mean velocity gradient of $0.1$~\kms~pc$^{-1}$, comparable to those of the clouds in the MW and Local Group galaxies, but smaller than those of the clouds in the ETGs studied so far (NGC~4429 and NGC~4526). These velocity gradients are likely induced by turbulence rather than large-scale galaxy rotation (\autoref{sec:kinematics}).
    
\item The GMCs of NGC~5806 have an unusually steep size -- linewidth relation ($\sigma_{\mathrm{obs,los}}\propto R_{\mathrm{c}}^{1.20\pm0.10}$; \autoref{fig:larson_only}), that may be due to gas inflows and shocks induced by the large-scale bar (\autoref{sec:discussion_steep_larson}).
    
\item The NGC~5806 GMCs are only marginally bound ($\langle\alpha_{\mathrm{vir}}\rangle\approx2$), and the virial parameters do not significantly differ across the different regions (see \autoref{fig:alpha_obs}). The virial parameters are positively correlated with the linewidths (see \autoref{fig:virial_relation}).

\item There are molecular gas inflows from the large-scale bar into the nuclear ring, with a velocity $V_{\rm in}\approx120$~\kms\ and a total mass inflow rate $\dot{M}_{\rm in}\approx5$~M$_{\odot}$~yr$^{-1}$ (\autoref{sec:discussion_turbulence}). These inflows could be at origin of the observed high velocity dispersions in the nuclear ring and the clouds therein.
    
\item The number of clouds decreases azimuthally from one node to the other within the nuclear ring, downstream from the nodes (\autoref{sec:discussion_gmc_ring}). By tracking cloud disruption through GMC number statistics, we estimate the typical cloud lifetime to be $\approx6$~Myr. This is larger than the estimated timescales of cloud-cloud collisions, shear and/or stellar feedback ($\approx3$~Myr), suggesting that any of those could contribute to the destruction of clouds within the nuclear ring.
\end{enumerate}

Overall, the large-scale bar seems to play an important role (via gas inflows and shocks) shaping the cloud population in the central region of NGC~5806, including potentially creating an unusually steep cloud size -- linewidth relation.

\section*{Acknowledgements}

We thank the anonymous referee for helpful and constructive comments and Prof.\ Daniel Wang for useful discussions. WC and AC acknowledge support by the National Research Foundation of Korea (NRF), grant Nos.\ 2018R1D1A1B07 048314, 2022R1A2C100298211 and 2022R1A6A1A03053472. LL was supported by a Hintze Fellowship, funded by the Hintze Family Charitable Foundation, and by a DAWN Fellowship, funded by the Danish National Research Foundation under grant No.\ 140. MB was supported by STFC consolidated grant ``Astrophysics at Oxford'' ST/H002456/1 and ST/K00106X/1. TAD acknowledges support from the UK Science and Technology Facilities Council through grants ST/S00033X/1 and ST/W000830/1. JG gratefully acknowledges financial support from the Swiss National Science Foundation (grant No. CRSII5 193826). This paper makes use of the following ALMA data: ADS/JAO.ALMA\#2016.1.00437.S and ADS/JAO.ALMA\#2016.2.00053.S. ALMA is a partnership of ESO (representing its member states), NSF (USA) and NINS (Japan), together with NRC (Canada), NSC and ASIAA (Taiwan) and KASI (Republic of Korea), in cooperation with the Republic of Chile. The Joint ALMA Observatory is operated by ESO, AUI/NRAO and NAOJ. This paper also makes use of observations made with the NASA/ESA \textit{Hubble Space Telescope}, obtained from the Hubble Legacy Archive, which is a collaboration between the Space Telescope Science Institute (STScI/NASA), the Space Telescope European Coordinating Facility (ST-ECF/ESA) and the Canadian Astronomy Data Centre (CADC/NRC/CSA). This research has made use of the NASA/IPAC Extragalactic Database (NED), which is operated by the Jet Propulsion Laboratory, California Institute of Technology, under contract with the National Aeronautics and Space Administration.

\section*{Data availability}

The data underlying this article are available in the ALMA archive (\url{https://almascience.eso.org/asax/}) under project code: (i) 2016.1.00437.S and (ii) 2016.2.00053.S. All analysed data will be shared upon request.


\bibliographystyle{mnras}
\bibliography{mnras_references}




\bsp	
\label{lastpage}
\end{document}